\documentclass[preprint,authoryear,11pt]{elsarticle}
\usepackage{amssymb,amsmath,array,graphicx,bm}
\usepackage{wrapfig}
\usepackage{amsthm}
\usepackage{float}
\restylefloat{table}
\usepackage{multirow}
\usepackage{xcolor} 
\usepackage{xr}
\usepackage[margin=1in]{geometry}
\usepackage{caption}
\usepackage{comment}
\usepackage{tikz}
\usetikzlibrary{shapes,arrows}
\usepackage{microtype} 
\usepackage[colorlinks=true,linkcolor=black, citecolor=blue, urlcolor=blue]{hyperref}
\usepackage{cases}
\usepackage{colortbl} 
\usepackage{array,booktabs}
\usepackage{babel,blindtext}
\usepackage{enumerate}

\theoremstyle{plain}

\usepackage[normalem]{ulem}

\tikzstyle{block} = [rectangle, draw,
    text width=7em, text centered, rounded corners, minimum height=4em]
\tikzstyle{line} = [draw, -latex']

\newlength{\tempheight}
\newlength{\tempwidth}

\newcommand*\patchAmsMathEnvironmentForLineno[1]{...} 
\newcommand*\patchBothAmsMathEnvironmentsForLineno[1]{...} 

\usepackage{caption}
\usepackage{subcaption}

\begin{document}
\begin{frontmatter}
\title{Comparative Analysis of Practical Identifiability Methods for an SEIR Model}

\author[vt]{Omar Saucedo}
\author[ttu]{Amanda Laubmeier}
\author[sdsu]{Tingting Tang}
\author[fsu]{Benjamin Levy}
\author[uiw]{Lale Asik}
\author[ox,uw]{Tim Pollington}
\author[utk]{Olivia Prosper}

\address[vt]{Department of Mathematics, Virginia Tech, VA 24061 }
\address[ttu]{Department of Mathematics and Statistics, Texas Tech University, TX 79409}
\address[sdsu]{Department of Mathematics and Statistics, San Diego State University, CA 92182}
\address[fsu]{Department of Mathematics, Fitchburg State University, MA 01420}
\address[uiw]{Department of Mathematics and Statistics, University of the Incarnate Word, TX 78209}
\address[ox]{ Big Data Institute, University of Oxford}
\address[uw] {Mathematics for Real-World System Center, University of Warwick, UK}
\address[utk] {Department of Mathematics, University of Tennessee, Knoxville, TN 37996}

\begin{abstract}
Identifiability of a mathematical model plays a crucial role in parameterization of the model. In this study, we establish the structural identifiability of a Susceptible-Exposed-Infected-Recovered (SEIR) model given different combinations of input data and investigate practical identifiability with respect to different observable data, data frequency, and noise distributions. The practical identifiability is explored by both Monte Carlo simulations and a Correlation Matrix approach. Our results show that practical identifiability benefits from higher data frequency and data from the peak of an outbreak. The incidence data gives the best practical identifiability results compared to prevalence and cumulative data. In addition, we compare and distinguish the practical identifiability by Monte Carlo simulations and a Correlation Matrix approach, providing insights into when to use which method for other applications.
\end{abstract}
\end{frontmatter}
\section{Introduction}
Compartment models have been used extensively to study infectious diseases. Among them, Susceptible-Exposed-Infectious-Recovered (SEIR) models have been extensively used to study disease dynamics impacted by vertical transmission (\cite{li2001global,gao2011seasonality}), vaccination strategies (\cite{d2002stability}), delayed infectiousness (\cite{yan2006seir}), multistage infectiousness (\cite{li2005global}),  treatment strategies (\cite{liu2019bifurcation}), and most recently, applied to COVID-19 (\cite{engbert2021sequential, efimov2021interval,ruktanonchai2020assessing}). These models often include parameters for which numerical values are unknown \emph{a priori} and cannot be directly measured. Since many parameters in a given model are not directly measurable, researchers often obtain parameters from outbreak data.

Parameter estimation relies on comparing empirical observations of the modeled system with the corresponding model output. Many computational techniques can be employed for parameter estimation. Most of these techniques rely on minimizing the difference between model output and observed data. However, before numerically estimating parameter values, it is important to address whether the parameters of the model are identifiable.
\emph{Identifiability analysis} determines to what extent and with what level of certainty the parameters of a model can be recovered from the available empirical data (\cite{borisov2020confidence}). These relate to two general types of identifiability analysis: structural and practical identifiability. \emph{Structural identifiability} is the theoretical possibility of determining the true values of parameters of a model from observations of its outputs and knowledge of its dynamic equations (\cite{bellman1970structural,cobelli1980parameter, walter1997identification}). On the other hand, \emph{practical identifiability} provides information on the accuracy with which parameters can be estimated from the available discrete and noise-corrupted measurements (\cite{martynenko2018intelligent}. So for an infinite amount of noise-free data, structural identifiability implies (the maximum) practical identifiability (\cite{wieland2021structural}).

Three common methods for evaluating the practical identifiability of models are Monte Carlo simulations (\cite{miao2011identifiability,tuncer2018structural}), correlation matrices (\cite{rodriguez2006hybrid,rodriguez2006novel,banks2014modeling}), and the profile likelihood (\cite{venzon1988method,jacquez1990parameter,raue2009structural,eisenberg2014determining,kao2018practical}). The \emph{Monte Carlo approach} can be applied to evaluate disparate model structures each with different observation schema by implementing a random sampling algorithm (\cite{miao2011identifiability}). While Monte Carlo simulations offer conceptual and algorithmic simplicity, their computational cost can be very high, as many samples are required to obtain a good approximation. The \emph{Correlation Matrix approach} assesses the correlation between parameter estimates (\cite{rodriguez2006hybrid,rodriguez2006novel,banks2014modeling}). It is much less computationally intensive than the Monte Carlo method but only provides a pairwise analysis of model parameters. Likelihood profiling is a common way to perform a practical identifiability analysis when using a likelihood-based estimation procedure (e.g. approximate Bayesian computation or maximum likelihood estimation) because the profiles are used for quantifying uncertainty in parameter estimates (i.e. to approximate confidence intervals) (\cite{raue2009structural}).

Epidemiological data plays a critical role in infectious disease surveillance, as it describes how infectious diseases are distributed within populations and what factors contribute to transmission (\cite{baron1996medical}). Several mathematical models have been established during the last decades to forecast disease progression using epidemiological data (\cite{lipsitch2003transmission,riley2003robust,fraser2009pandemic,tuite2011cholera,chowell2014transmission,fisman2014early}). There are many technical challenges in implementing a standardized epidemiological data framework. The way data is reported and collected, and type and frequency of data available through data-sharing institutions (public health departments, ministries of health, data collection, or aggregation services) differ significantly. For example, different data types (including prevalence, incidence, cumulative incidence, and mortality) are reported at differing time intervals (such as daily, weekly, monthly, or even more infrequently). As a result, understanding how these factors affect parameter identifiability and estimation is critical.

In this study, we systematically examined the influence of different data types (prevalence, incidence, and cumulative incidence) and sampling frequencies (daily, weekly, and monthly) on the practical identifiability of SEIR model parameters. We explored four scenarios with different peak infection times and data collection windows, to assess the impact of different data collection strategies on parameter identifiability. Discrepancies between Monte Carlo and Correlation Matrix results highlight the importance of considering multiple criteria for identifiability. Overall, our results show that incidence data yields the most identifiable parameters, prevalence data exhibited intermediate identifiability, and cumulative incidence data resulted in the least identifiable parameters.  Varying data collection frequencies affected parameter identifiability, with longer time series and higher sampling rates generally improving identifiability. 

In Section \ref{sec:model}, we present the SEIR model and the structural identifiability results.  Section \ref{MCsimulation} and \ref{CM method} outline the use of MC and CM methods to assess parameter identifiability respectively. In Section \ref{sec:Results}, we present the results of our analysis, with Section \ref{MCResults} detailing the outcomes of the MC simulations, Section \ref{sec:CMResults} illustrating the CM results, and Section \ref{sec:Comparison} compare the results from both MC and CM methods. In Section \ref{sec:Discussion}, we discuss our finding, their implications, and limitations. Finally, Section \ref{sec:conclusion}, we provide a concise summary of our findings.

\section{The SEIR model}
\label{sec:model}
\begin{figure}[htbp]
\centering
 \tikzstyle{block} = [rectangle, draw, fill=gray!40,
    text width=2em, text centered, rounded corners, minimum height=3em, minimum width=3em]
  \begin{tikzpicture}[node distance=2.0cm, auto, >=stealth]
   \coordinate(S) at (0,0);
   	\node[block] (S)             {\Large$S$};

    \node [] (Birth) [left of=S, node distance=2cm]{};

	\node[block](E)[right of = S, node distance=2cm] 	{\Large$E$};

    \node[block](I)[right of=E, node distance=2cm] 	{\Large$I$};

    \node[block](R)[right of = I, node distance=2cm] 	{\Large$R$};

    \node [] (Smort) [left of=S, above of = S, node distance=1cm]{};

    \node [] (Emort) [left of=E, above of = E, node distance=1cm]{};

    \node [] (Imort) [left of=I, above of = I, node distance=1cm]{};

    \node [] (Idismort) [right of=I, below of = I, node distance=1cm]{};

    \node [] (Rmort) [left of=R, above of = R, node distance=1cm]{};

    \draw[->](S)--node[pos=.5,left,sloped,anchor=center,above]{$\beta I$}(E);

    \draw[->](E)--node[pos=.5,left,sloped,anchor=center,above]{$\gamma$}(I);	

    \draw[->](I)--node[pos=.5,left,sloped,anchor=center,above]{$\alpha$}(R);
 \end{tikzpicture}
 \caption {Flowchart for general SEIR framework.}
 \label{fig:model1naive}
\end{figure}
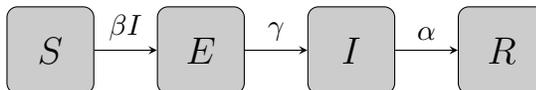

The overarching goal of our study is to examine how different data types and collection frequencies influence parameter identifiability under structural and practical identifiability paradigms. To this end, we chose a simple SEIR model framework containing a small number of parameters as the foundation for our study.  Here, the SEIR model is a closed population with no vital dynamics (i.e. no births or deaths):
\begin{align}
\label{eqn:system}
    \dot{S}(t) &= -\beta S I, \nonumber \\
    \dot{E}(t)&= \beta S I - \gamma E, \nonumber\\
    \dot{I}(t)&= \gamma E - \alpha I, \\
    \dot{R}(t) &= \alpha I \nonumber, \\
    \dot{C}(t) &= \beta S I \nonumber,
\end{align}
where $N(t)=S(t)+E(t)+I(t)+R(t)$ is the total population, the parameter $\beta $ is the transmission rate from susceptible to infected, $\gamma$ is the rate of transition from exposed to infectious, and $\alpha $ is the recovery rate from infected to recovered (Figure \ref{fig:model1naive}). The state variable $I(t)$ represents prevalence of the infectious disease at time $t$.  The auxiliary variable $C(t)$ tracks the cumulative number of infectious individuals from the start of the outbreak. $C(t)$ is not a state of the system but rather a class to track cumulative incidence. The number of new infections at time $t$ is $\dot{C}(t)$, but in practice, incidence of new cases is observed as $C(t)-C(t-1)$. Initial conditions for the $S,E,I,$ and $R$ states will be noted by $S(0),E(0),I(0),$ and $R(0)$, respectively.

The SEIR model Eq.\eqref{eqn:system} is well studied and often presented as an entry-level educational tool (\cite{brauer2008mathematical}). As individuals progress through the compartments, eventually the number of susceptible individuals may be insufficient to sustain transmission. This leads to the Disease-Free Equilibrium, which is globally asymptotically stable when $\beta N /\alpha < 1$ and unstable when $\beta N /\alpha >1$.
This cutoff is the basic reproduction number $\mathcal{R}_0$, which is commonly used to characterize the infectiousness of a disease that is freshly introduced into a completely susceptible population.
Importantly, this reproduction number $\mathcal{R}_0$, along with other epidemiological metrics, are calculated from model parameters. If a model is non-identifiable, then researchers risk reaching incorrect conclusions about disease persistence that follow from incorrect parameter values.

\subsection{Structural identifiability}
\label{sec:StructuralIdent}
\noindent  \emph{Structural identifiability} addresses the question of whether the parameters of a model can be uniquely determined from `perfect' input-output data, assuming the underlying model correctly represents the input-output data. Here, perfect data means the data is continuous in time and without noise. Consider a general compartment model:
\begin{equation}\label{dynamicalmodel}
\begin{aligned}
      \dot{\mathbf{x}}(t) &= f\big(\mathbf{}{x}(t),\mathbf{p}\big), \\
      \mathbf{x}(0) &= \mathbf{x}_0, \\
      \mathbf{y}(t) &= \mathbf{g}\big(\mathbf{x}(t),\mathbf{p}\big),
\end{aligned}
\end{equation}
where the $n$ state variables at time $t$ are represented by the state vector $\mathbf{x}(t)\in\mathbb{R}^n$, parameters are denoted by the vector $\mathbf{p}\in\mathbb{R}^k$, the initial condition vector is denoted as $\mathbf{x}_0\in\mathbb{R}^n$, and the $m$ observation variables are represented by $\mathbf{y}(t)\in\mathbb{R}^m$.
\noindent The system Eq.\eqref{dynamicalmodel} is \emph{globally structurally identifiable} for the parameter set $\mathbf{p}_1 $ if for every parameter set $\mathbf{p}_2,$  $$y(t,\mathbf{p}_1)=y(t,\mathbf{p}_2)\implies \mathbf{p}_1=\mathbf{p}_2.$$

\noindent We say that system Eq.\eqref{dynamicalmodel} is \emph{locally structurally identifiability} for the parameter set $\mathbf{p}_1$ if for every $\mathbf{p}_2$, $$y(t,\mathbf{p}_1)=y(t,\mathbf{p}_2), \text{ and } \mathbf{p}_2\in B(\mathbf{p}_1)\implies \mathbf{p}_1=\mathbf{p}_2, \text{ where } B(\mathbf{p}_1) \text{ is a ball centered at } \mathbf{p}_1.$$

\noindent Otherwise, system Eq.\eqref{dynamicalmodel} is \emph{unidentifiable}. An equivalent characterization of \emph{local identifiability} is if there are at least two or more finite sets of $\mathbf{p}$ that yield the same observations $\mathbf{y}$. Here, the state variables denoted by $\mathbf{x}$ are specifically $[x_1,x_2,x_3,x_4,x_5]=[S,E,I,R,C].$

There are several well-established methods for determining the structural identifiability of disease models such as the direct test (\cite{denis2001some,walter2004guaranteed}) and the similarity transformation method (\cite{vajda1989similarity, evans2002identifiability,yates2009structural}), which can be applied to autonomous (no external input) systems. Alternatively, the Taylor series approach (\cite{pohjanpalo1978system}), differential algebra (\cite{bellu2007daisy,ljung1994global}), the generating series approach (\cite{walter1981unidentifiable}), implicit functions approach (\cite{xia2003identifiability}), or differential geometry (\cite{villaverde2016structural}) are applicable
to external input. Various software tools have been developed to perform structural identifiability analysis: \texttt{Observability Test} (\cite{sedoglavic2002probabilistic}), Differential Algebra for Identifiability of SYstems (\texttt{DAISY}, \cite{bellu2007daisy}), Exact Arithmetic Rank (\texttt{EAR}, \cite{anguelova2012minimal}), \texttt{COMBOS} (\cite{meshkat2014finding}), Data2Dynamics (\cite{raue2015data2dynamics}), \texttt{STRIKE-GOLDD} (\cite{villaverde2016structural}), \texttt{GenSSI2} (\cite{ligon2018genssi}), and Structural Identifiability Analyser (\texttt{SIAN}, \cite{hong2019sian}). These toolboxes provide a broad range of features and implement methods based on differential algebra, semi-numerical differential algebra, generating series, local algebraic observability, and identifiability tables. However, some of these toolboxes have limited performance or can be time-consuming to use (\cite{rey2023benchmarking}).

We establish the structural identifiability of the prevalence and cumulative incidence data types using the software packages \texttt{DAISY} and \texttt{SIAN}. As far as we know, all of these packages can only handle observations when explicitly expressed as functions of the state variables at any time. Consequently, we are unable to examine the structural identifiability of the SEIR model for incidence defined as a difference in time, so its analysis will be limited to practical identifiability. \texttt{DAISY} provides algebraic relationships among parameters in a form to find parameter combinations. In contrast, \texttt{SIAN} uses a combination of differential algebra methods with the Taylor series approach. It constructs a mapping that binds the parameter values and initial conditions to the output functions. By replacing these output functions with truncated versions of their Taylor series, the mapping is transformed into another mapping between finite-dimensional spaces. To achieve this, the order of truncation is determined so that it contains enough information for the identifiability analysis. The accuracy of the result is represented by an estimated probability.

Besides the difference in the underlying theoretical framework for both packages, there are two key distinctions in application. Firstly, \texttt{SIAN} does not allow an explicit definition of the initial conditions; instead, it treats them as parameters. Secondly, \texttt{DAISY} outputs explicit algebraic expressions for model parameters and how unidentifiable parameters relate algebraically, while \texttt{SIAN} does not. Therefore, we employ both \texttt{SIAN} and \texttt{DAISY} for assessing the structural identifiability of the model when initial conditions are unknown, while exclusively relying on \texttt{DAISY} for scenarios with specified initial conditions. Table \ref{tab:structural ident result} summarizes the outcomes of these assessments. In cases with specified initial conditions, all parameters of the SEIR model exhibit structural identifiability concerning both prevalence and cumulative incidence outputs. However, when initial conditions are unspecified, only $\beta$ is structurally identifiable with respect to prevalence, while the remaining parameters are only locally identifiable. Cumulative incidence shows all parameters are only locally identifiable.

\begin{table}[H]
\centering
 \begin{tabular}{m{50mm} m{30mm} m{30mm} }
  \midrule
  \bfseries Observables & \bfseries Globally-Identifiable &\bfseries Locally- Identifiable \\
  \midrule  \midrule
   \bfseries Prevalence (I) & $\beta$ &$\gamma,\alpha $  \\
   \cmidrule(l r){1-3}
   \bfseries Prevalence (I with ICs) & $ \beta,\gamma,\alpha$ \\
   \cmidrule(l r){1-3}
    \bfseries Cumulative (C) &  &$\beta,\gamma,\alpha$ \\
       \cmidrule(l r){1-3}
    \bfseries Cumulative (C with ICs) &  $\beta,\gamma,\alpha$ &\\
   \bottomrule
\end{tabular}
\caption{Structural identifiability of $\beta,\gamma,\alpha$ with or without initial conditions (ICs) for different observables. The results for the cases without ICs using both \texttt{SIAN} and \texttt{DAISY} were identical.}
\label{tab:structural ident result}
 \end{table}

\section{Methods for practical identifiability}
\label{PracticalIdent}
Compared to structural identifiability analysis, practical identifiability analysis accounts for the sampling rates and noisiness of experimental data (\cite{lizarralde2020sensitivity}). For this reason, practical identifiability analyses typically involve fitting models to epidemic data. In general, a model is considered \textit{practically identifiable} if a unique parameter set for a given model can be consistently obtained. Researchers may use real or simulated epidemic data to assess practical identifiability (\cite{wu2008parameter,eisenberg2013identifiability,roosa2019assessing,tuncer2018structural}), and there are different approaches and criteria for assessing identifiability.
In this study, we apply two approaches to analyze the practical identifiability of an SEIR model with hypothetical `true' parameters. We first use a Monte Carlo approach to consider the practical identifiability of our system of ODEs before conducting a similar analysis using a Correlation Matrix approach.

In general, practical identifiability can be defined for the dynamical model in Eq.\eqref{dynamicalmodel}. In this study, we consider $\mathbf{x}=[S, E, I, R, C]$ and $f(\mathbf{x}(t),\mathbf{p})$ the right hand side of Eq.\eqref{eqn:system}.  We assume that $\mathbf{x}_0$ is known but that all parameters $\mathbf{p}=[\beta, \gamma, \alpha]$ must be recovered from data. For practical identifiability, observations $\mathbf{y}(t)$ at a finite set of time points $t_i$ are modeled by: \begin{equation}\label{observationmodel}
\begin{aligned}
       {\mathbf{y}(t_i)} &= g\big( \mathbf{x}(t_i),\mathbf{p} \big) \big( 1+\epsilon(t_i) \big),
\end{aligned}
\end{equation} with random measurement noise $\epsilon(t_i)$ drawn from a normal distribution with zero mean and standard deviation $\sigma$, i.e., $\epsilon \sim \mathcal{N}(0,\sigma)$. The set of time points $t_i$ are sampled at different frequencies and time spans depending on the scenario. In this study, we denote $\mathbf{y}$ as prevalence, incidence, or cumulative incidence data and define $g\big(\mathbf{x}(t),\mathbf{p}\big)$ to select the appropriate state: $g(\mathbf{x}(t_i),\mathbf{p}) = x_3(t_i)$ for prevalence, $g\big(\mathbf{x}(t_i),\mathbf{p} \big)=x_5(t_i)$ for cumulative incidence, and $g\big(\mathbf{x}(t_i),\mathbf{p} \big) = x_5(t_i)-x_5(t_{i-1}) $ for incidence. A model is practically identifiable if there is a unique parameter set $\mathbf{p} = \hat{\mathbf{p}}$ that minimizes the difference between the model output $g$ and the data $\mathbf{y}$.

Data collection plays a key role when estimating parameters for an epidemic model. For example, case data may be collected on a daily, weekly, or monthly basis.  To determine how various time series data impact the practical identifiability of the model parameters, we explored four scenarios with different peaks of the epidemic curve and lengths of data collection.
See Supplemental Figure \ref{fig:scenarios} for plots of each scenario.
 \begin{enumerate}[\indent {}]
    \item \textbf{\textit{Scenario 1:}} The peak for the epidemic curve occurred at day 109 with the time span of 365 days, where $\beta=0.0001$, $\gamma=0.2$, and  $\alpha=0.03$.
    \item \textbf{\textit{Scenario 2:}} The peak for the epidemic curve occurred at day 25 with the time span of 50 days, where $\beta=0.001$, $\gamma=0.2$, and  $\alpha=0.03$.
    \item \textbf{\textit{Scenario 3:}} The peak for the epidemic curve occurred at day 109 with the time span of 100 days.
    \item \textbf{\textit{Scenario 4:}} The peak for the epidemic curve occurred at day 25 with the time span of 20 days.
\end{enumerate}

 Note that the time spans for Scenarios 1 and 2 were chosen {to include roughly twice the time it takes for incidence (new infections) to drop below one. We selected this cutoff as a balance between capturing the entire prevalence curve and avoiding many observations near steady state for incidence and cumulative incidence.} Scenarios 3 and 4 simulate the effect of only having access to data before the peak of the epidemic.

\subsection{Monte Carlo simulations}\label{MCsimulation}
The history of Monte Carlo (MC) simulations can be traced back to the work of \cite{metropolis1949monte}. MC method, as a sampling technique using random numbers and probability distributions, can be used to determine the practical identifiability of a model. This approach is due to its versatility and straightforward implementation.
We perform MC simulations by generating $M=10,000$ synthetic data sets using the true parameter set  $\hat{\mathbf{p}}$ and adding noise to the data in increasing amounts. MC simulations are outlined in the following steps:
\begin{enumerate}
\item Solve the SEIR ODEs (Eq.\ref{eqn:system}) numerically with the true parameter vector $\hat{\mathbf{p}}$ to obtain the output vector $\mathbf{g}\big(\mathbf{x}(t),\hat{\mathbf{p}}\big)$ at the discrete data time points $\{t_i\}_{i=1}^n$ .
\item Generate $M=10,000$ data sets with a given measurement error. We assume the error follows a normal distribution with mean $0$ and variance $\sigma^2(t)$; that is, the data are described by $\mathbf{y}_{i,j} = g\big(\mathbf{x}(t_i),\hat{\mathbf{p}}\big)(1+\epsilon_{i,j})$, where $\epsilon \sim \mathcal{N}(0,\sigma)$ at the discrete data time points $\{t_i\}_{i=1}^n$ for all for $j=1,2,...,M$.
\item Estimate the parameter set $\mathbf{p}_j$, by fitting the dynamical model to each of the $M$ simulated data sets. This is achieved by minimizing the difference between model output and the data generated for the specific scenario:
\[
\mathbf{p}_j\approx\min_{\mathbf{p}} \sum_{i=1}^{n} \dfrac{\displaystyle \big(\mathbf{y}_{i,j}-g(\mathbf{x}(t_i),{\mathbf{p}})\big)^{2}}{(g(\mathbf{x}(t_i),{\mathbf{p}}))^2}.
\]
This optimization problem is solved in MATLAB R2021a using the built-in function \texttt{fminsearchbnd}, which is part of the Optimization Toolbox. Since \texttt{fminsearchbnd} is a local solver, the optimized minimum value can be influenced by the starting point. To avoid issues related to the starting value, we use the true parameter values as the initial parameter starting point provided to \texttt{fminsearchbnd}. Furthermore, we used the optimization function \texttt{fmincon} to test the consistency of the results in the methodology. Both functions produced the same qualitative results with similar AREs, the detailed ARE results are provided in the Supplemental document.  For the remainder of the manuscript, the results shown refer to AREs from  \texttt{fminsearchbnd}.
\item Calculate the average relative estimation error (ARE) for each parameter in the set $\mathbf{p}$ following \cite{miao2011identifiability}:
    \begin{equation}
       \mathrm{ARE} \big(p^{(k)}\big)=100 \%  \times \frac{1}{M}  \sum_{j=1}^{M} \frac{\left| \hat{p}^{(k)}- p_j^{(k)}\right|}{\left| \hat{p}^{(k)}\right| },
       \label{ARE_eqn}
    \end{equation}
    where $p^{(k)}$ is the $k$-th parameter in the set $\mathbf{p}$, $\hat{p}^{(k)}$ is the $k$-th parameter in the true parameter set $\hat{\mathbf{p}}$, and $ p_j^{(k)}$ is the $k$-th element of $\mathbf{p}_j$.
\item Repeat steps 1 through 4, increasing the level of noise ($\sigma = 0, 1, 5, 10, 20,  30\%)$.
\end{enumerate}
Following the convention in \cite{tuncer2018structural}, we assume that a given parameter is practically identifiable if the ARE of the parameter is less than or equal to the measurement error, $\sigma$.

\subsection{Correlation Matrix method}
\label{CM method}
Although the MC simulation approach is easy to understand and simple to implement, the associated computational cost is high due to a large number of repetitions and the associated optimization costs. One alternative is to utilize the sensitivity matrix of the model to compute the coefficient matrix for parameters of interest (\cite{banks2010standard,banks2014modeling,jacquez1985numerical}). This requires much less computation and is relatively simple if measurement errors follow an identical and independent distribution. The Correlation Matrix (CM) method assesses the correlation between estimated parameters using a matrix of output sensitivities to model parameters. If estimated parameters are highly correlated, then they are considered practically unidentifiable.

CM assessments of identifiability are local to a particular parameter set. The assessment also depends on the type of model observation, frequency of data collection, and assumed distribution of measurement noise. However, CM does not incorporate realizations of noisy observations, as in the MC approach. Instead, these considerations are incorporated in the following steps:
\begin{enumerate}
    \item Solve the SEIR model sensitivities numerically, for model observations $\mathbf{g}(\mathbf{x}(t),\mathbf{p})$ at the discrete collection times $\{t_i\}_{i=1}^n$ with respect to each parameter in $\mathbf{p}$.
    \begin{enumerate}
    \item For prevalence and cumulative incidence, this is obtained from the SEIR model sensitivity equations and extract solutions for $x_3$ ($I$) and $x_5$ ($C$): \begin{equation*}
        \frac{d}{dt}\frac{\partial \mathbf{x}}{\partial \mathbf{p}} = \frac{\partial f}{\partial \mathbf{x}}\frac{\partial \mathbf{x}}{\partial\mathbf{p}} + \frac{\partial f}{\partial \mathbf{p}}.    \end{equation*}
    \item For incidence, this is obtained from numerically integrating the quantity \begin{equation*} \frac{\partial}{\partial \mathbf{p}}\int_{t_i}^{t_{i+1}}\beta S I dt =  \int_{t_i}^{t_{i+1}} \left[\beta S \frac{\partial I}{\partial \mathbf{p}} + \left(\beta \frac{\partial S}{\partial \mathbf{p}} + \frac{\partial \beta}{\partial \mathbf{p}} S\right) I \right]dt,\end{equation*} where $\frac{\partial I}{\partial \mathbf{p}}$ and $\frac{\partial S}{\partial \mathbf{p}}$ come from solutions to the sensitivity equations.
    \end{enumerate}
    \item Construct a sensitivity matrix where the $i,j^{\text{th}}$ component corresponds to the sensitivity of the model output $\mathbf{g}(\mathbf{x}(t),\mathbf{p})$ at time $t_i$ to the $j^{\text{th}}$ parameter in $\mathbf{p}$. This is denoted by \begin{equation*} F_{i,j}=\frac{\partial \mathbf{g}}{\partial p_j}(\mathbf{x}(t_i),\mathbf{p}).\end{equation*}
    \item Compute the inverse of the weighted Fisher Information Matrix, $IM=(F^TWF)^{-1}$, where $W$ is the diagonal weighted matrix for least squares error. For the assumed distribution of error in Eq.\eqref{observationmodel}, the weights are $g(\mathbf{x}(t_i),\mathbf{p})$.
    \item Compute the correlation coefficients $\chi_{ij}=IM_{ij}/\sqrt{IM_{ii}IM_{jj}}$ for all $i,j=1,2,3$ and $i\neq j$, corresponding to correlations between the three parameters in $\mathbf{p}$.
    \item If the correlation coefficients are below 0.9 for all parameter pairs, then the parameter set is practically identifiable for the assumed model observations.\end{enumerate}

We conduct the CM approach for all data types described at the start of Section \ref{PracticalIdent}, to assess the practical identifiability of the model for the true parameter sets. We also repeat this process for estimated parameters ${\mathbf{p}}_j$ obtained in Step 3 of the MC approach (Section \ref{MCsimulation}). This allows us to assess the perceived identifiability of parameter estimates obtained from noisy data.

\section{Results}
\label{sec:Results}

In this section, we will examine the output of the MC and CM methods described in Section \ref{PracticalIdent} when applied to the SEIR model described in Section \ref{sec:model}. For all the estimated parameters, we chose the search bounds in the \verb|fminsearchbnd| algorithm to be $[0,1]$. Although in practice we may not know the true bounds of the parameters, we chose the biologically possible range of the parameters with respect to their epidemiological definitions. We are interested in how different factors impact practical identifiability results and whether these results are aligned for both methods (MC and CM). We do not expect to have a one-to-one relationship in each scenario; however, it is beneficial to examine when and why we obtain matching outcomes when they occur in the simulations.

\subsection{Monte Carlo (MC) results}
\label{MCResults}
 Using the MC algorithm in Section \ref{MCsimulation}, we tested the practical identifiability of the SEIR model parameters with respect to different data types, sampling frequencies, and time period of available data. A model parameter is said to be \emph{practically identifiable} if the ARE values given by Eq.\eqref{ARE_eqn} are less than or equal to the noise percentage $\sigma_0$ for the given noise level. For example, if the ARE values for all levels of noise for $\beta$ are less than the corresponding noise level, we say $\beta$ is practically identifiable. If the ARE for $\beta$ is more than its corresponding noise level for any noise level, we would say this parameter is not practically identifiable. If all parameters are practically identifiable in a given scenario, then the model is practically identifiable. This is consistent with the definition of practical identifiability used in \cite{tuncer2018structural}. The identifiability results from the MC approach are summarized in Table \ref{table:MC ARE}, and the details of the ARE values can be found in the Supplemental Table \ref{Prevalence Scenario 1 FMSB}-\ref{Cumulative Incidence Scenario 4 FMSB}. 

\begin{figure}[h!]
    \centering
    \begin{subfigure}[b]{0.32\textwidth}{\includegraphics[width=\textwidth]{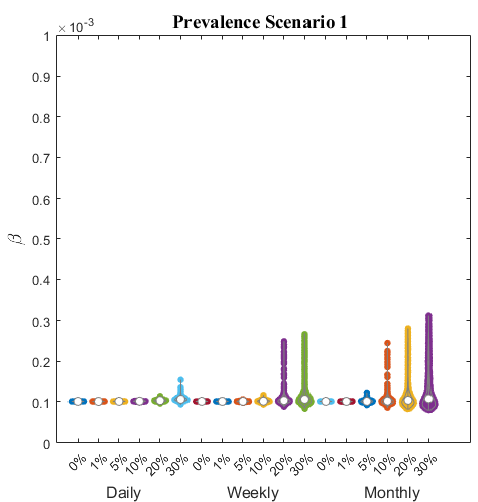}}\end{subfigure}
     \begin{subfigure}[b]{0.32\textwidth}{\includegraphics[width=\textwidth]{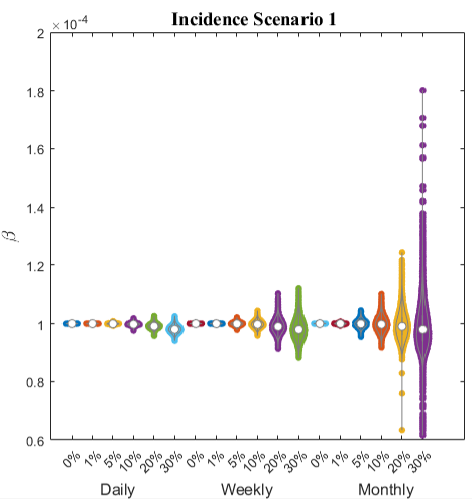}}\end{subfigure}
        \begin{subfigure}[b]{0.32\textwidth}{\includegraphics[width=\textwidth]{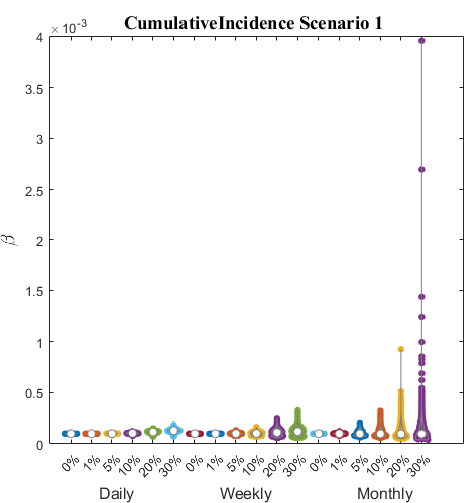}}\end{subfigure}
    \caption{Violin Plots for Prevalence, Incidence, and Cumulative Incidence in Scenario 1.}
    \label{fig:ViolinScenario1}
\end{figure}

\begin{figure}[H]
    \centering
    \includegraphics[scale=.35]{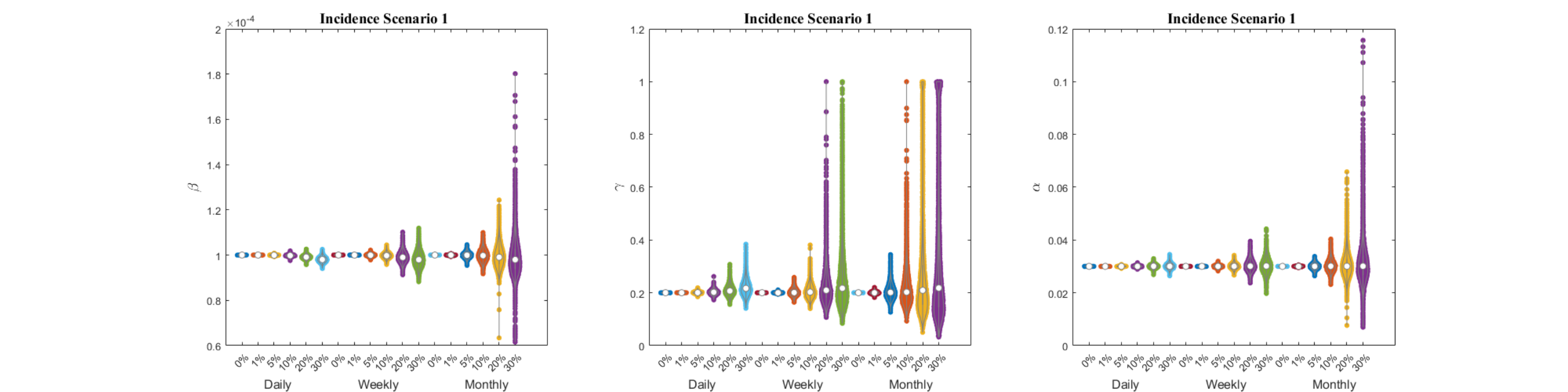} \\
      \includegraphics[scale=.35]{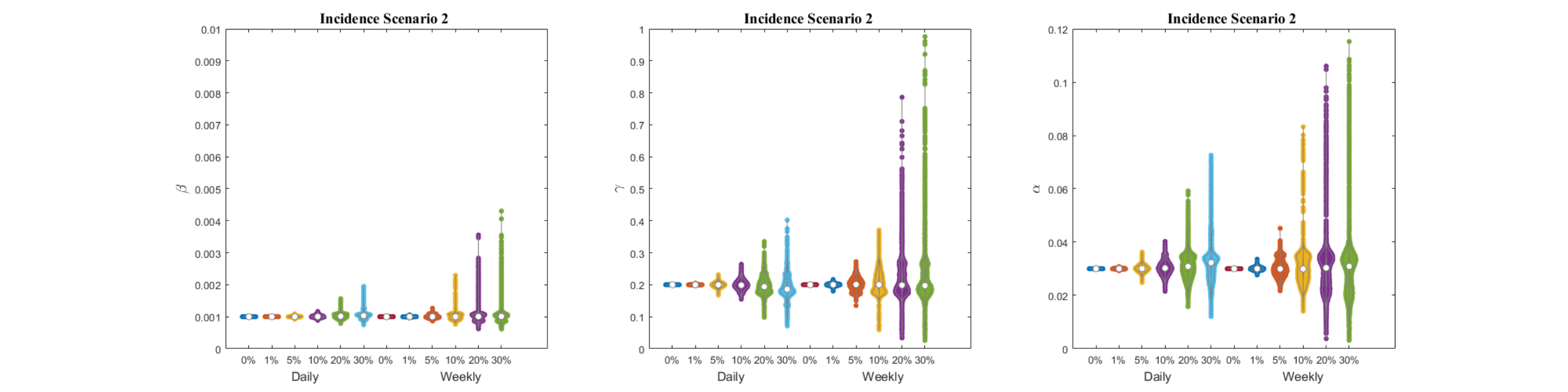} \\
        \includegraphics[scale=.35]{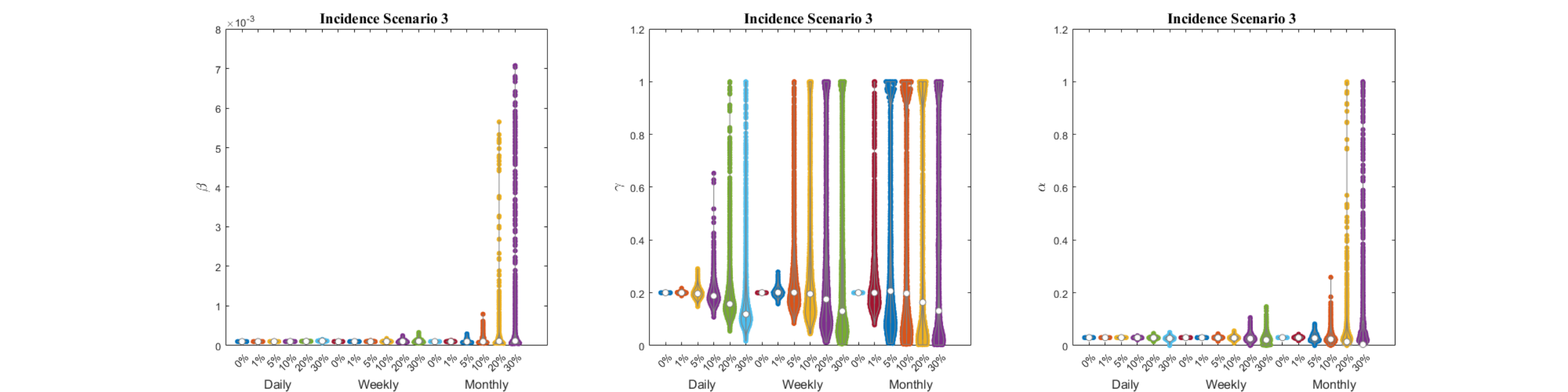} \\
          \includegraphics[scale=.35]{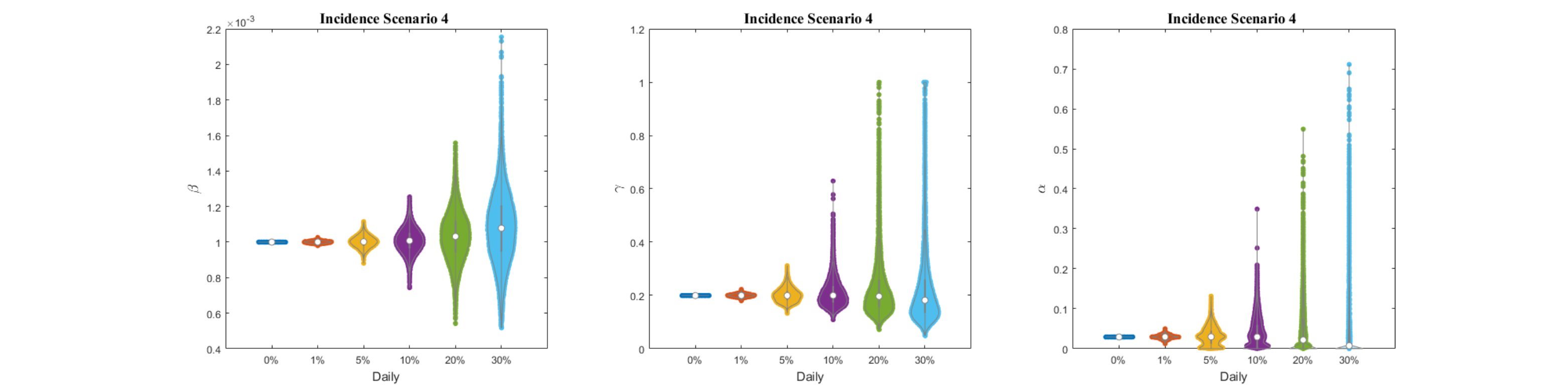} \\
    \caption{Violin Plots for $\beta$, $\gamma$, and $\alpha$ using Incidence data for all four scenarios. These distributions are generated from MC algorithm using 10,000 iterations.}
    \label{fig:ViolinScenariosIncid}
\end{figure}

\begin{figure}[H]
    \centering
    \includegraphics[scale=.35]{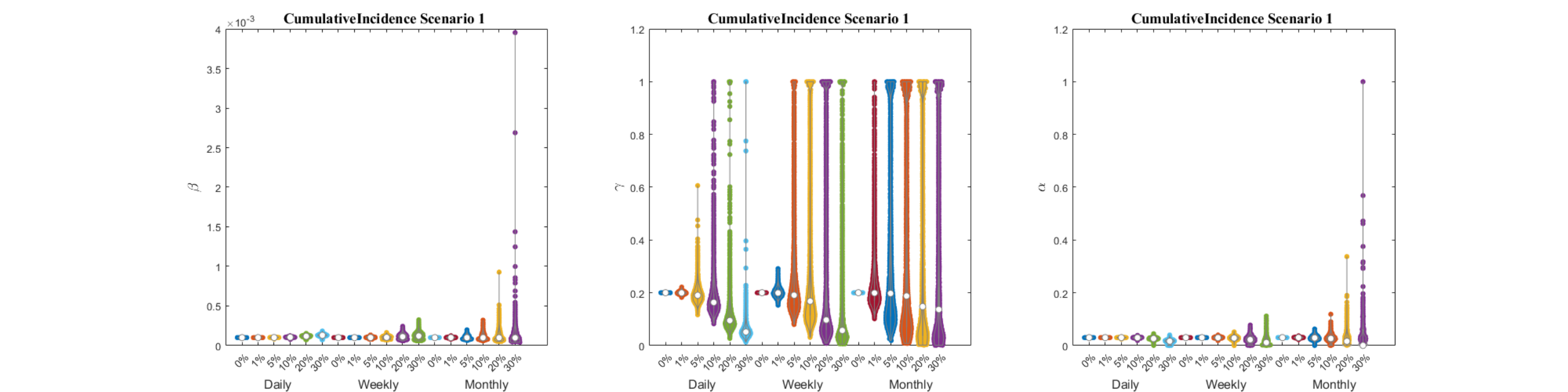} \\
      \includegraphics[scale=.35]{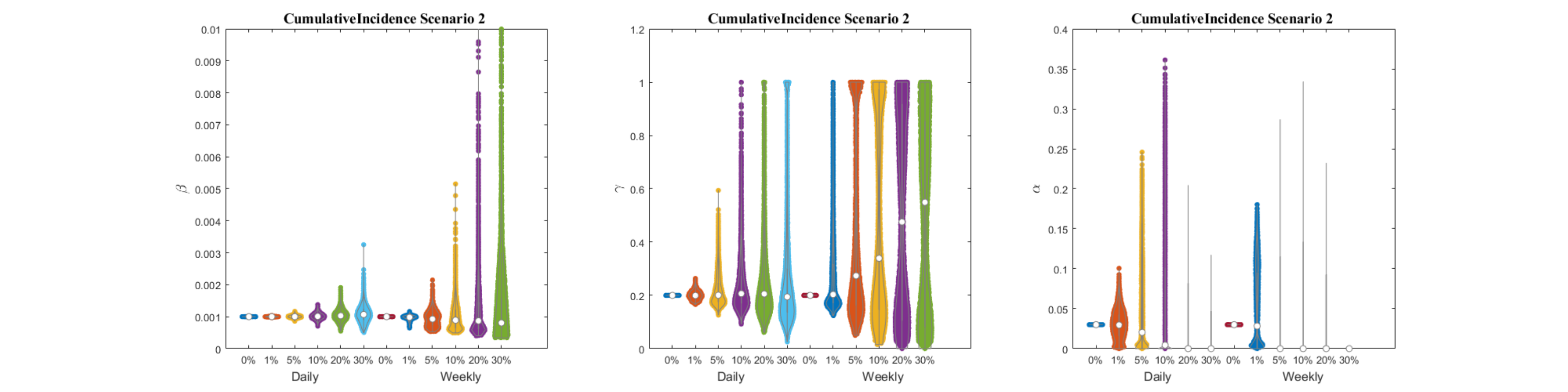} \\
        \includegraphics[scale=.35]{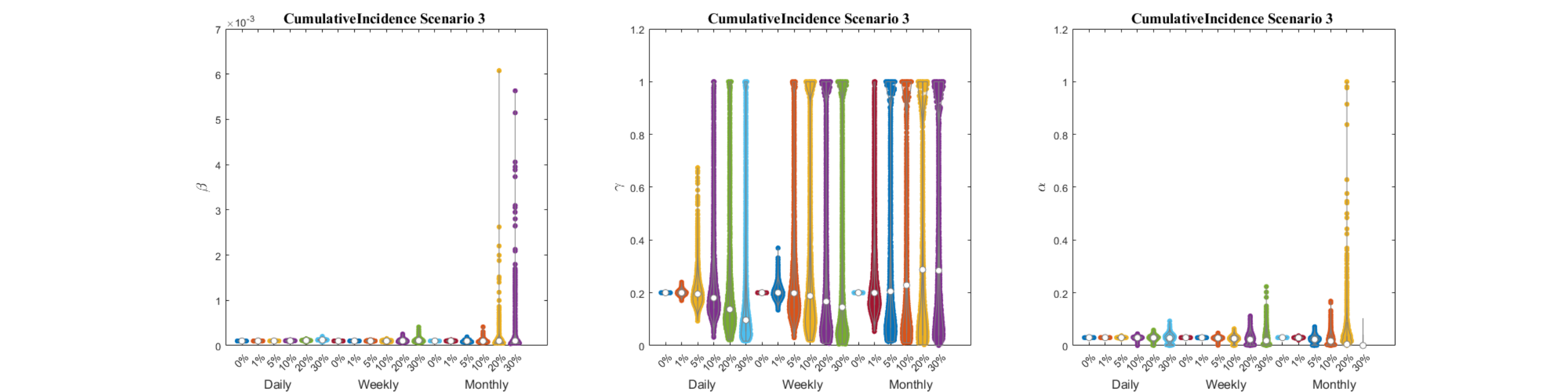} \\
          \includegraphics[scale=.35]{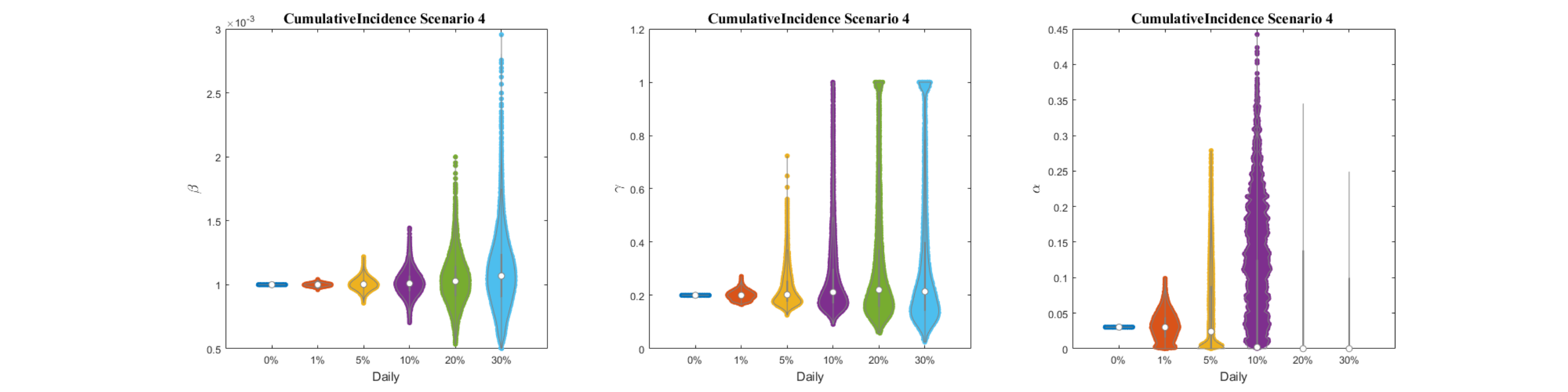} \\
    \caption{Violin Plots for $\beta$, $\gamma$, and $\alpha$ using Cumulative Incidence data for all four scenarios. These distributions are generated from MC algorithm using 10,000 iterations.}
    \label{fig:ViolinScenariosCumul}
\end{figure}

\begin{figure}[H]
    \centering
    \includegraphics[width=0.7\textwidth]{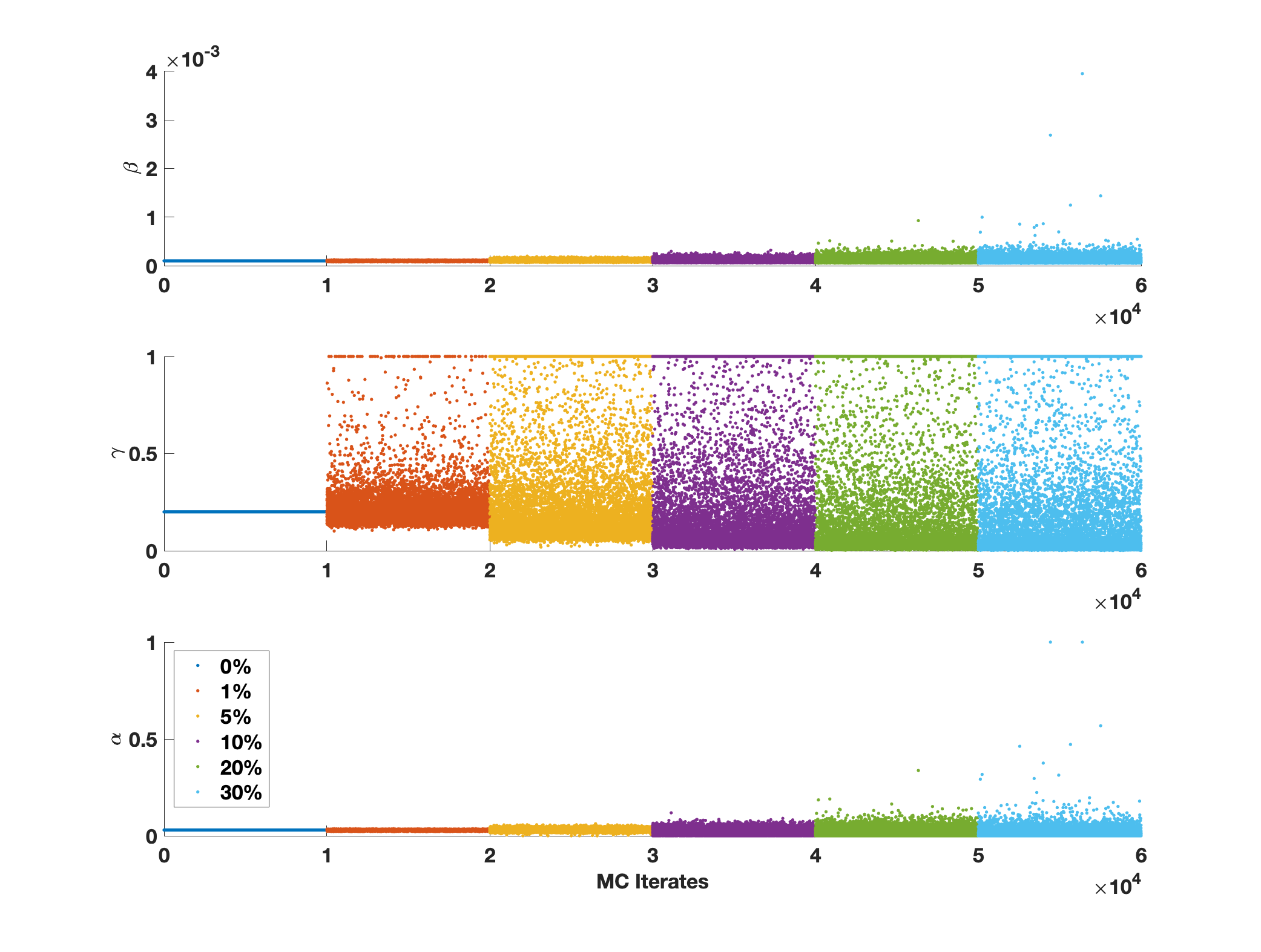}
    \caption{Raw MC parameter estimates (10,000) for the data Cumulative Incidence Scenario 1 Monthly.}
    \label{fig:MCAllParams}
\end{figure}

Before examining the ARE values, we generate violin plots for each scenario to observe the distribution of parameter values as noise is introduced in the MC procedure. Although we are not using the violin plots as a condition for practical identifiability, it does provide insight into how the values for each parameter change as we incorporate noise.  Figure \ref{fig:ViolinScenario1} represents the violin plots for prevalence, incidence, and cumulative incidence in Scenario 1 with respect to $\beta$. Predictably, as more noise is added to the data set, we observe a larger dispersion of parameter values for $\beta$, especially for 20\% and 30\% noise levels.  The remaining plots for the other scenarios, data type, and parameters can be found in the Supplemental materials Figure \ref{fig:ViolinScenariosPrev}, \ref{fig:ViolinScenariosIncid}, \ref{fig:ViolinScenariosCumul}, and \ref{fig:MCAllParams}.


\begin{table}[h!]
\centering
\begin{tabular}{c|ccc|ccc|ccc}
Prevalence &               \multicolumn{3}{c|}{Daily}                          &                                 \multicolumn{3}{c|}{Weekly}                             &                               \multicolumn{3}{c}{Monthly}                       \\ \hline
Scenario 1          & \cellcolor[HTML]{8EA9DB}$\beta$ & \cellcolor[HTML]{8EA9DB}$\gamma$ & \cellcolor[HTML]{8EA9DB}$\alpha$ & \cellcolor[HTML]{8EA9DB}$\beta$ & $\gamma$ & \cellcolor[HTML]{8EA9DB}$\alpha$ & \cellcolor[HTML]{8EA9DB}$\beta$ & $\gamma$ & \cellcolor[HTML]{8EA9DB}$\alpha$  \\
Scenario 2          & $\beta$                         & $\gamma$                         & \cellcolor[HTML]{8EA9DB}$\alpha$ & $\beta$                         & $\gamma$ & \cellcolor[HTML]{8EA9DB}$\alpha$ &                         &  &                     \\
Scenario 3          & \cellcolor[HTML]{8EA9DB}$\beta$ & $\gamma$                         & \cellcolor[HTML]{8EA9DB}$\alpha$ & $\beta$                         & $\gamma$ & $\alpha$                         & $\beta$                         & $\gamma$ & $\alpha$                    \\
Scenario 4          & $\beta$                         & $\gamma$                         & $\alpha$                         &                              &       &                               &                              &       &
\end{tabular}

\vspace{.5cm}

\centering
\begin{tabular}{c|ccc|ccc|ccc}
Incidence  &               \multicolumn{3}{c|}{Daily}                          &                                 \multicolumn{3}{c|}{Weekly}                             &                               \multicolumn{3}{c}{Monthly}                    \\ \hline
Scenario 1         & \cellcolor[HTML]{8EA9DB}$\beta$ & \cellcolor[HTML]{8EA9DB}$\gamma$ & \cellcolor[HTML]{8EA9DB}$\alpha$ & \cellcolor[HTML]{8EA9DB}$\beta$ & $\gamma$                         & \cellcolor[HTML]{8EA9DB}$\alpha$ & \cellcolor[HTML]{8EA9DB}$\beta$ & $\gamma$ & \cellcolor[HTML]{8EA9DB}$\alpha$  \\
Scenario 2         & \cellcolor[HTML]{8EA9DB}$\beta$ & \cellcolor[HTML]{8EA9DB}$\gamma$ & \cellcolor[HTML]{8EA9DB}$\alpha$ & \cellcolor[HTML]{8EA9DB}$\beta$ & \cellcolor[HTML]{8EA9DB}$\gamma$ & \cellcolor[HTML]{8EA9DB}$\alpha$ &                         &  &                         \\
Scenario 3         & \cellcolor[HTML]{8EA9DB}$\beta$ & $\gamma$                         & \cellcolor[HTML]{8EA9DB}$\alpha$ & \cellcolor[HTML]{8EA9DB}$\beta$                         & $\gamma$                         & $\alpha$                         & $\beta$                         & $\gamma$ & $\alpha$                        \\
Scenario 4         & \cellcolor[HTML]{8EA9DB}$\beta$                         & $\gamma$                         & $\alpha$                         &                              &                               &                               &                              &       &
\end{tabular}

\vspace{.5cm}

\begin{tabular}{c|ccc|ccc|ccc}
Cumulative &                              &                             &                          &                              &                          &                        &                              &       &                    \\
Incidence &               \multicolumn{3}{c|}{Daily}                          &                                 \multicolumn{3}{c|}{Weekly}                             &                               \multicolumn{3}{c}{Monthly}           \\ \hline
Scenario 1         &  \cellcolor[HTML]{8EA9DB}$\beta$ &  $\gamma$ &  $\alpha$ &  $\beta$ & $\gamma$                         &  $\alpha$ &  $\beta$ & $\gamma$ & $\alpha$ \\
Scenario 2         &  \cellcolor[HTML]{8EA9DB}$\beta$ &  $\gamma$ &  $\alpha$ &  $\beta$ &  $\gamma$ &  $\alpha$ &                      &  &                         \\
Scenario 3         &  \cellcolor[HTML]{8EA9DB}$\beta$ & $\gamma$                         & $\alpha$ & \cellcolor[HTML]{8EA9DB}$\beta$                         & $\gamma$                         & $\alpha$                         & $\beta$                         & $\gamma$ & $\alpha$                         \\
Scenario 4         & \cellcolor[HTML]{8EA9DB}$\beta$                         & $\gamma$                         & $\alpha$                         &                              &                               &                               &                              &       &
\end{tabular}
\caption{Practically identifiability results for all scenarios: prevalence, incidence, and cumulative incidence were using \emph{fminsearchbnd} and \emph{fmincon} with 10,000 iterations. Both methods yield the same identifiability results.   }
\label{table:MC ARE}
\end{table}
\par As expected, a higher data density and longer temporal data availability tend to result in greater parameter identifiability overall. Moreover, the parameter $\gamma$ had the fewest identifiable individual scenarios, with the highest number of identifiable scenarios being $\beta$ and $\alpha$ depending on the disease data metric used. More scenarios of $\alpha$ were identifiable when using prevalence, while incidence resulted in more scenarios of $\beta$ being identifiable. Additionally, the incidence scenarios had the most cases where parameters were identifiable, while only $\beta$ is identifiable in some cases with cumulative incidence data.  In particular, one special case where shorter temporal data yields better identifiability results occurred for parameter $\beta$ with cumulative incidence data. This may be attributed to the performance of the optimization algorithm. It is also worth noting that comparing the weekly (monthly) results  from Scenario 1 and 2 with the daily (weekly) results from Scenario 3 and 4, one could conclude that knowing more about the whole epidemic curve is more important for obtaining identifiable model than obtaining dense data for a shorter period of time.  A detailed discussion and reasoning about these findings are presented in Section~\ref{sec:Discussion}.
\subsection{Correlation Matrix (CM) results}
\label{sec:CMResults}

We next test the practical identifiability of the same scenarios as in the MC results, using the CM approach.  Since CM reports identifiability of parameter pairs, instead of individual parameters, we holistically define a problem to be identifiable if all three parameter pairs have a correlation below 0.9. If any pair is more correlated, then the remaining estimates may be affected and result in an unidentifiable problem. In Supplemental Tables \ref{CMPrev}-\ref{CMCuml}, we report these correlation results and in Table \ref{table:CMApproach}, we summarize the resulting assessments of practical identifiability. We find that the most identifiable data type is incidence data, which leads to identifiable problems for any sampling rate in Scenario 1 and daily sampling rates in Scenarios 3 and 4. Estimating from prevalence data is an identifiable problem for daily and weekly sampling rates in Scenario 3, and problems using cumulative incidence are never identifiable.

In most cases, parameter correlation values do not significantly change with sampling frequency. Only six correlation values increase by more than 0.1 between daily and monthly data, and none of these changes cross the 0.9 threshold for identifiability. Four correlation values increase past the 0.9 threshold between daily and monthly data, changing the assessment of identifiability for two data types (monthly prevalence and incidence data in Scenario 3). We therefore find two cases (out of nine) where increased sampling frequency improves the CM assessment of identifiability. However, this is a limited number of occurrences and it does not suggest an overall relationship between sampling frequency and identifiability under the CM criteria. In all other cases, the CM results are not sensitive to frequency of data collection. Contrary to expectations, four correlation values decrease between daily and monthly data, although the change is less than 0.1 and does not cross the 0.9 threshold for identifiability.

\begin{table}[]
\centering
\begin{tabular}{c|ccc|ccc|}
  & \multicolumn{3}{c|}{\bfseries Prevalence}                                 & \multicolumn{3}{c|}{\bfseries Incidence}                                                           \\

  & \bfseries Daily                           & \bfseries Weekly                           &\bfseries Monthly  &\bfseries Daily                           &\bfseries Weekly                           &\bfseries Monthly                           \\
    \hline
\bfseries Scenario 1 & No                          & No                          & No & \cellcolor[HTML]{8EA9DB}Yes & \cellcolor[HTML]{8EA9DB}Yes & \cellcolor[HTML]{8EA9DB}Yes \\
\bfseries Scenario 2 & No                          & No                          &    & No                          & No                          &                             \\
\bfseries Scenario 3 & \cellcolor[HTML]{8EA9DB}Yes & \cellcolor[HTML]{8EA9DB}Yes & No & \cellcolor[HTML]{8EA9DB}Yes & No                          & No                          \\
\bfseries Scenario 4 & No                          &                             &    & \cellcolor[HTML]{8EA9DB}Yes &                             &
\end{tabular}
\caption{Practical identifiability for the CM approach, for different scenarios and data types. Cumulative incidence is not shown, since all results were `not identifiable.'}
\label{table:CMApproach}
\end{table}

\subsection{Comparing MC and CM}  
\label{sec:Comparison}

In Supplemental Tables \ref{MCMCPrev}-\ref{MCMCCuml}, we assess identifiability using the CM criteria for all estimates obtained in the MC process, under $0\%$ and $30\%$ noise levels, then report the percentage of MC estimates that would be considered identifiable. At $0\%$ noise level, the CM results for the MC estimates match results for the true parameters, since the estimates are very close to the true parameters. However, at $30\%$ noise level, the MC estimates may be far from the true parameter set. For example, using daily prevalence data, the true parameter set for Scenario 3 is identifiable by the CM approach. However, only $18\%$ of the MC parameter estimates at $30\%$ noise are identifiable by the CM criteria. Although $\beta$ and $\alpha$ are identifiable by the MC criteria and remain close to their true values, there is a higher error in the estimate for $\gamma$. Because the CM results are affected by these incorrect $\gamma$ values, the entire problem is considered unidentifiable. In other cases, incorrect MC estimates may be considered identifiable by the CM criteria. For example, using daily or weekly cumulative incidence data, the true parameter set for Scenario 1 is not identifiable by the CM criteria. Under the MC criteria, $\beta$ is identifiable using daily data, but very close to the cutoff, and all other parameters are not identifiable. Using weekly data, none of the parameters are identifiable. However, $66\%$ of the MC parameter estimates using daily data and $34\%$ of the MC estimates using weekly data are identifiable by the CM criteria.

We next consider parameter correlations in four different cases, which correspond to all possible assessments of identifiability compared across the MC and CM criteria. The cases include daily prevalence data from Scenario 1, weekly prevalence data from Scenario 3, daily incidence data from Scenario 1, and daily cumulative incidence data from Scenario 1.  In Supplemental Figure \ref{fig:ScatterPlotMCParameterPairs}, we plot the normalized error of MC estimates using data with $30\%$ noise, for estimated pairs of $\beta$:$\gamma$, $\beta$:$\alpha$, and $\alpha$:$\gamma$. We fit a straight line through the parameter error and report the slope, as a numerical assessment of correlation between MC parameter estimates. We consider two parameters correlated if the lines are sufficiently different from a strictly horizontal or vertical fit, which we define to be when the magnitude of the slope is between 0.5 and 2. Using this threshold, the two cases identifiable under the MC criteria also have uncorrelated parameter estimates. We compare these values to the CM results and present the correlations and identifiability in Table \ref{correlations}. We find that the sign of the correlation matches across the MC estimates and CM calculations in all but one case ($\alpha$:$\gamma$ for daily cumulative incidence data from Scenario 1). However, there is not a consistent match in the correlation magnitudes across MC estimates and CM calculations. The overall classification of `correlated' matches for nine (out of 12) pairs, but the thresholds for both criteria are somewhat arbitrary.

\begin{table}[H]
\centering
\begin{tabular}{|l|cc|llllll|}
\hline
\multicolumn{1}{|c|}{\textbf{Scenario}}      & \multicolumn{2}{c|}{\textbf{Identifiability}}                                                       & \multicolumn{6}{c|}{\textbf{Correlations}}                                                                                                                                                                      \\ \hline
\rowcolor[HTML]{FFFFFF}
                                             & \multicolumn{1}{c|}{\cellcolor[HTML]{FFFFFF}MC} & \multicolumn{1}{c|}{\cellcolor[HTML]{FFFFFF}CM} & \multicolumn{3}{c|}{\cellcolor[HTML]{FFFFFF}MC}                                                                & \multicolumn{3}{c|}{\cellcolor[HTML]{FFFFFF}CM}                                              \\
                                             & \multicolumn{1}{l|}{}                             &                                                 & $\beta$:$\gamma$              & $\beta$:$\alpha$             & \multicolumn{1}{l|}{$\alpha$:$\gamma$}            & $\beta$:$\gamma$              & $\beta$:$\alpha$             & $\alpha$:$\gamma$             \\ \hline
\rowcolor[HTML]{8EA9DB}
\cellcolor[HTML]{FFFFFF}S1 Prevalence daily  & \multicolumn{1}{l|}{\cellcolor[HTML]{FFFFFF}yes}  & \cellcolor[HTML]{FFFFFF}no                      & -2.86                         & -0.22                        & \multicolumn{1}{l|}{\cellcolor[HTML]{8EA9DB}6.12} & \cellcolor[HTML]{FFFFFF}-0.98 & -0.66                        & 0.54                          \\
\rowcolor[HTML]{8EA9DB}
\cellcolor[HTML]{FFFFFF}S3 Prevalence weekly & \multicolumn{1}{l|}{\cellcolor[HTML]{FFFFFF}no}   & \cellcolor[HTML]{FFFFFF}yes                     & {\cellcolor[HTML]{FFFFFF}-1.17}                        & {\cellcolor[HTML]{8EA9DB}0.15}                       & {\cellcolor[HTML]{FFFFFF}-1.27} & -0.87                         & 0.89                         & -0.57                         \\
\rowcolor[HTML]{8EA9DB}
\cellcolor[HTML]{FFFFFF}S1 Incidence daily   & \multicolumn{1}{l|}{\cellcolor[HTML]{FFFFFF}yes}  & \cellcolor[HTML]{FFFFFF}yes                      & -9.77                         & -0.20                        & \multicolumn{1}{l|}{\cellcolor[HTML]{8EA9DB}2.51} & -0.84                         & -0.14                        & 0.64                          \\
\cellcolor[HTML]{FFFFFF}S1 Cumulative daily  & \multicolumn{1}{l|}{\cellcolor[HTML]{FFFFFF}no}   & \cellcolor[HTML]{FFFFFF}no                      & \cellcolor[HTML]{FFFFFF}-1.83 & 0.38                         & \multicolumn{1}{l|}{\cellcolor[HTML]{8EA9DB}0.10} & \cellcolor[HTML]{FFFFFF}-0.97 & \cellcolor[HTML]{FFFFFF}0.92 & \cellcolor[HTML]{8EA9DB}-0.81 \\ \hline
\end{tabular}
\caption{Parameter correlations for four cases, corresponding to different assessments of identifiability. For the MC approach, the `correlation' is the slope of a best-fit line between parameter estimates. For the CM approach, the correlation comes from calculations in Section \ref{sec:CMResults} A blue-shaded cell indicates that the two parameters are classified as `not correlated.'}
\label{correlations}
\end{table}

\section{Discussion}
\label{sec:Discussion}

In this study, we systematically investigated the impact of different data types (prevalence, incidence, and cumulative incidence) and sampling frequencies (daily, weekly, and monthly) on the practical identifiability of the SEIR model parameters, using two identifiability methods: the Monte Carlo (MC) method and the Correlation Matrix (CM) method. We found that incidence data, sampled at a higher frequency, resulted in the greatest degree of identifiability. While more data (as obtained at a higher sampling frequency) should intuitively lead to greater identifiability (although some exceptions are discussed in more detail below), it is less clear why incidence data should lead to greater identifiability compared to cumulative data. As such, we examine more closely the structure of these data types, and how they relate to model parameters. The disease metrics incidence and cumulative incidence are related to the flow into the exposed compartment, $E$, which depends on the transmission rate parameter $\beta$. On the other hand, prevalence is the result of integrating over the flows into and out of the infectious class, $I$, which are directly impacted by disease-progression rate $\gamma$ and recovery rate $\alpha$, and only indirectly by $\beta$. Because cumulative incidence, `smooths out',  the information encoded in incidence, it is possible that different combinations of disease progression and recovery rates can yield very similar cumulative incidence curves. For example, if a decrease in the incubation period causes an increased rate of flow into the infected class, this could be balanced out by an increase in the recovery rate out of the infected class and thus lead to similar values of the cumulative incidence at the current time. This hypothesis is further supported by the transmission rate being the only identifiable parameter for cumulative incidence. 

A more significant contributor, however, to the unidentifiability of cumulative incidence data, is how cumulative incidence is generated. We argue that the method for adding measurement error in the case of cumulative data is patently unrealistic. The error model presented in our work (motivated by other studies on identifiability in outbreak models) assumes that the mean of the measurement error is proportional to the solution curve $C(t)$, resulting in very large measurement errors late in the epidemic. A more realistic way to generate the bootstrapped cumulative data would be to add measurement error to the incidence data and accumulate the incidence to obtain the corresponding cumulative incidence data, which results in much smaller measurement errors later in the epidemic compared with the first approach described here -- a result of incidence tapering off after the peak of the epidemic. The challenge with the second approach is that the accumulated incidence data produces measurement errors that are not independent across time.

 There are identifiable parameters in each of the various levels of data collection frequency, but the same cannot be said for each scenario. That is, given the low enough noise in our data, we are able to identify at least one parameter for each frequency of data collection, but we find that supplying data for only a portion of an epidemic is a limiting factor for practical identifiability. This is evident in Scenarios 3 and 4, where truncated time series data is used in the estimation process, resulting in few, if any, practically identifiable parameters. Since these scenarios mimic what would occur during an outbreak, care should be taken when estimating parameters before the epidemic has reached its peak, as there is likely to be a great deal of uncertainty in the estimates, and therefore, in the trajectories that follow. Projections of hypothetical control problems should take this uncertainty into consideration.

 There are more identifiable problems by the CM criteria than problems where all parameters are identifiable by the MC criteria. However, CM does not allow for partially identifiable problems, and there are more cases where the MC approach classifies some parameters as independently identifiable than fully identifiable problems under the CM approach. The MC and CM assessments of `completely identifiable' problems seldom match, except for Scenario 1 daily incidence data. The assessments do, however, agree on the `completely unidentifiable' problems for Scenarios 1 and 2 using weekly or monthly cumulative incidence data, Scenario 3 monthly data of all types, and Scenario 4 daily prevalence data. For both assessments, we find that problems using incidence data are most often identifiable and problems using cumulative incidence data are least often identifiable. The assessments do not appear to follow similar patterns across sampling rates or cutoff dates. The MC approach more often follows the pattern that increased sampling rates (daily vs monthly) or longer time series (Scenario 3 vs Scenario 1) lead to more identifiable problems, while CM is less sensitive to sampling rates and is inconsistent in its response to longer time series.

Overall, we find that more data does correspond to more identifiable problems for the MC approach or does not affect identifiable problems for the CM approach. However contrary to expectations, we find some cases where recovering parameters using shorter time series (Scenarios 3 or 4) are identifiable problems, where problems using longer time series (Scenarios 1 or 2) were not identifiable. Under the MC criteria when using weekly cumulative incidence data, $\beta$ is identifiable for Scenario 3 but not Scenario 1. This also occurs under the CM criteria using daily incidence data, where Scenario 4 is an identifiable problem but Scenario 2 is not. These results suggest that long tails without change in the time series may reduce identifiability. For the MC approach, the long tails may present an opportunity for the minimization to `fit to noise'. This is supported by the difference across error levels, where the estimates have more error at lower noise levels for Scenario 3 (Supplemental Table \ref{Cumulative Incidence Scenario 3 FMSB}) but as noise increases, the estimates for Scenario 1 have more error (Supplemental Table \ref{Cumulative Incidence Scenario 1 FMSB}). Intuitively this also makes sense for the CM approach, which relies on the model's sensitivity matrix; once the model solution reaches zero or levels off, there is less sensitivity to the model parameters. This explanation does not hold for prevalence data, which does not level off in our scenarios, but we find that under the CM criteria when using daily or weekly prevalence data, Scenario 3 is an identifiable problem but Scenario 1 is not. We do not have an explanation for this outcome, except that Scenario 3 is borderline unidentifiable (0.89 correlation, in Supplemental Table \ref{CMPrev}) and that the 0.9 cutoff for identifiability may not be appropriate.

In Section \ref{sec:Comparison}, we further examine how MC and CM assessments of identifiability are related. In general, we found little overlap between the two, and by applying the CM criteria to MC estimates, we find that the CM criteria can lead to misleading results about parameter estimates. Most importantly, we found that parameter estimates that did not meet MC criteria for identifiability were sometimes identifiable by the CM criteria. This indicates that in practice, it is possible to estimate incorrect values for unidentifiable parameters but conclude by the CM criteria that the recovered parameters were identifiable. In other cases, the true parameter set was identifiable by the CM criteria and the MC remained close to the true parameters, but the MC estimates were unidentifiable by the CM criteria. This discrepancy would also carry over to sensitivity-based confidence intervals, making it possible to have low confidence in a well-estimated parameter due to relatively small changes in the parameter space. However, it is important to note that our MC results represent a best case, in which numerical minimization starts from the true parameter estimate; in practice, the high sensitivities that affect the CM approach may affect estimates that start from other parts of the parameter space. In assessing the correlations of MC estimates, we find that the sign of correlation between parameters matched the results from the CM approach, but there is not a clear relationship between the magnitude of correlation for both approaches.

For the MC method, we tested two optimizers, \texttt{fmincon} and \texttt{fminsearchbnd}, from MATLAB for numerically finding the optimal parameter sets. \texttt{fmincon} is a function typically used for constrained optimization where the parameter space is limited through equality and inequality constraints and the objective function is at least second-order differentiable.  It uses an interior point algorithm to find the optimal solution by default unless otherwise specified. On the other hand, \texttt{fminsearchbnd} requires upper and lower bounds for the parameter spaces, but can't handle any other constraints on the parameters. It uses the Nelder-Mead simplex direct search algorithm which doesn't require differentiability of the objective function. We implemented these two functions not to directly compare the numerical results, but to determine if they provided similar identifiability outputs.

We computed the AREs six times with different optimization algorithms, number of iterations, and data generation processes for each of the 27 cases (9 cases for each of the three data types as shown in Table~\ref{table:MC ARE}) and presented the results for the run with the most iterations using \texttt{fminsearchbnd} and \texttt{fmincon} in the Supplemental Material. Notably, these computations exhibit both qualitative and quantitative variations. The first two times, we generate 1,000 simulated data for each case, employing the \texttt{fmincon} and \texttt{fminsearchbnd} optimization methods separately. These two optimizers give different identifiability for 5 cases. To mitigate the stochastic variability arising from data generation, we computed the AREs two additional times using the same 1,000 simulated data for each case across both optimizers. Then, the number of different cases is only one, though a different case than the previous 5. Furthermore, to reduce inter-case stochastic variability resulting from the utilization of partial data from Scenarios 1 and 2 in Scenarios 3 and 4, we conduct the final two rounds of computations. In these rounds, we initially generated 10,000 simulated data sets for Scenarios 1 and 2 and subsequently extracted the required data sets for all other cases. We observe that both optimizers yielded consistent qualitative outcomes across all cases. Based on this observation, we recommend increasing the number of simulated data sets whenever feasible, in order to enhance the reliability of the results.

Quantitatively, there are cases where AREs fluctuate around the $30\%$ threshold over these six times. For example, the ARE corresponding to $30\%$ noise data for $\beta$ with prevalence as observation ranges from $28.3\%$ to $30.9\%$. It would be difficult to assert practical identifiability in such cases. There are arguments that practical identifiability can be assessed as long as these thresholds are relatively on the same level, however, we found there is one case where the AREs ranges from $16\%$ to $63\%$ meanwhile another case the AREs only varies from $16.1\%$ to $17.1\%$. This poses a difficulty in the certainty of the results from the MC method. Our intensive computations suggest that the typically used 1,000 simulations for MC method may be insufficient in practice.

 In addition, two distinct computational behaviors emerge when computing AREs for the same data:  1) in some cases when the parameters are not identifiable, \texttt{fmincon} produces notably larger ARE than \texttt{fminsearchbnd;} 2) Furthermore, there are occurrences where the ARE computed from  \texttt{fminsearchbnd} is nearly zero while the results from \texttt{fmincon}  distinctly deviate from zero. The first phenomenon is likely due to the utilization of an unbounded search region by the \texttt{fmincon} optimizer, which arises from the absence of constraints on the parameter space. In contrast, when employing the \texttt{fminsearchbnd} optimizer, we impose a range of $[0,1]$ for all parameters, thereby obtaining less AREs by confining the search within a bounded parameter space. The second phenomenon could be attributed to the search algorithm employed by these two optimizers. Our investigation indicates that in instances where a non-zero ARE is obtained for zero noise data, the \texttt{fmincon} optimizer generates optimal parameter values that deviate from the accurate ones, even when the initial parameter values are set to the accurate values. Notably, the resulting parameters yield a larger error compared to the accurate values. This suggests that the algorithm encounters random deviations from the initial search point and is unable to converge back to the accurate parameters, potentially due to the intricate landscape of the error function. \\

\section{Conclusion}
\label{sec:conclusion}

In this study, we found that for a range of parameter values, sampling rates, and criteria for practical identifiability, parameters estimated from incidence data are most often identifiable and parameters estimated from cumulative incidence data are least often identifiable. In practice, estimates obtained from cumulative incidence data should be treated with caution. However, it is important to note that identifiability was sensitive to underlying parameters across different scenarios. Additionally, we found that assessments of identifiability seldom agreed across the MC and CM approaches, with each method relying on cutoffs for identifiability that may be arbitrary. A further complication is that parameter estimation and identifiability results were sensitive to the choice of minimization algorithm, even after implementing common safeguards through numerical tolerances. Taken together, sensitivities to parameter values, definition of identifiability, and numerical algorithms demonstrate that a single assessment of identifiability may be misleading or incomplete. We recommend a range of tests for identifiability to ensure confidence in results. Finally, we note that our results are obtained for a relatively simple epidemic model under a best-case scenario, in which the model perfectly matches the processes generating the data, and were generated with knowledge of the underlying parameter set, allowing for numerical searches to begin from the desired solutions. These conditions are impossible to meet in practice, and we still found fewer identifiable problems than expected. Parameter estimation for more complex models must be treated with significant caution and may require more rigorous testing of identifiability than is currently practiced. Although challenging, work in this area would greatly benefit from concrete guidelines for assessing practical identifiability from an array of methods. \\

\noindent {\bf Acknowledgements} This paper is based on work supported by the American Mathematical Society Mathematics Research Communities through National Science Foundation grants 1641020 and 1916439. OP was supported in part by National Science Foundation grant DMS-2045843.

\bibliographystyle{spbasic}
\bibliography{Identifiability}

\begin{thebibliography}{58}
\providecommand{\natexlab}[1]{#1}
\providecommand{\url}[1]{{#1}}
\providecommand{\urlprefix}{URL }
\expandafter\ifx\csname urlstyle\endcsname\relax
  \providecommand{\doi}[1]{DOI~\discretionary{}{}{}#1}\else
  \providecommand{\doi}{DOI~\discretionary{}{}{}\begingroup
  \urlstyle{rm}\Url}\fi
\providecommand{\eprint}[2][]{\url{#2}}

\bibitem[{Anguelova et~al.(2012)Anguelova, Karlsson, and
  Jirstrand}]{anguelova2012minimal}
Anguelova M, Karlsson J, Jirstrand M (2012) Minimal output sets for
  identifiability. Mathematical Biosciences 239(1):139--153,
  \doi{10.1016/j.mbs.2012.04.005}

\bibitem[{Banks et~al.(2010)Banks, Holm, and Robbins}]{banks2010standard}
Banks HT, Holm K, Robbins D (2010) {Standard Error Computations for Uncertainty
  Quantification in Inverse Problems: Asymptotic Theory vs. Bootstrapping}.
  {Mathematical and Computer Modelling} 52(9-10):1610--1625,
  \doi{10.1016/j.mcm.2010.06.026}

\bibitem[{Banks et~al.(2014)Banks, Hu, and Thompson}]{banks2014modeling}
Banks HT, Hu S, Thompson WC (2014) {Modeling and Inverse Problems in the
  Presence of Uncertainty}. CRC Press, \doi{10.1201/b16760}

\bibitem[{Baron(1996)}]{baron1996medical}
Baron S (1996) {Medical Microbiology}

\bibitem[{Bellman and {\AA}str{\"o}m(1970)}]{bellman1970structural}
Bellman R, {\AA}str{\"o}m KJ (1970) {On Structural Identifiability}.
  Mathematical Biosciences 7(3-4):329--339, \doi{10.1016/0025-5564(70)90132-X}

\bibitem[{Bellu et~al.(2007)Bellu, Saccomani, Audoly, and
  D’Angi{\`o}}]{bellu2007daisy}
Bellu G, Saccomani MP, Audoly S, D’Angi{\`o} L (2007) {DAISY: a new software
  tool to test global identifiability of biological and physiological systems}.
  Computer Methods and Programs in Biomedicine 88(1):52--61,
  \doi{10.1016/j.cmpb.2007.07.002}

\bibitem[{Borisov and Metelkin(2020)}]{borisov2020confidence}
Borisov I, Metelkin E (2020) Confidence intervals by constrained
  optimization—an algorithm and software package for practical
  identifiability analysis in systems biology. PLOS Computational Biology
  16(12):e1008,495

\bibitem[{Brauer et~al.(2008)Brauer, Van~den Driessche, Wu, and
  Allen}]{brauer2008mathematical}
Brauer F, Van~den Driessche P, Wu J, Allen LJ (2008) Mathematical Epidemiology,
  vol 1945. Springer, \doi{10.1007/978-3-540-78911-6}

\bibitem[{Chowell and Nishiura(2014)}]{chowell2014transmission}
Chowell G, Nishiura H (2014) {Transmission dynamics and control of Ebola virus
  disease (EVD): a review}. BMC Medicine 12(1):1--17,
  \doi{10.1186/s12916-014-0196-0}

\bibitem[{Cobelli and DiStefano~3rd(1980)}]{cobelli1980parameter}
Cobelli C, DiStefano~3rd JJ (1980) Parameter and structural identifiability
  concepts and ambiguities: a critical review and analysis. American Journal of
  Physiology-Regulatory, Integrative and Comparative Physiology 239(1):R7--R24,
  \doi{10.1152/ajpregu.1980.239.1.R7}

\bibitem[{Denis-Vidal et~al.(2001)Denis-Vidal, Joly-Blanchard, and
  Noiret}]{denis2001some}
Denis-Vidal L, Joly-Blanchard G, Noiret C (2001) Some effective approaches to
  check the identifiability of uncontrolled nonlinear systems. Mathematics and
  computers in simulation 57(1-2):35--44, \doi{10.1016/S0378-4754(01)00274-9}

\bibitem[{d'Onofrio(2002)}]{d2002stability}
d'Onofrio A (2002) {Stability properties of pulse vaccination strategy in SEIR
  epidemic model}. Mathematical Biosciences 179(1):57--72,
  \doi{10.1016/s0025-5564(02)00095-0}

\bibitem[{Efimov and Ushirobira(2021)}]{efimov2021interval}
Efimov D, Ushirobira R (2021) {On an interval prediction of COVID-19
  development based on a SEIR epidemic model}. Annual Reviews in Control
  51:477--487, \doi{10.1016/j.arcontrol.2021.01.006}

\bibitem[{Eisenberg and Hayashi(2014)}]{eisenberg2014determining}
Eisenberg MC, Hayashi MA (2014) Determining identifiable parameter combinations
  using subset profiling. Mathematical Biosciences 256:116--126,
  \doi{10.1016/j.mbs.2014.08.008}

\bibitem[{Eisenberg et~al.(2013)Eisenberg, Robertson, and
  Tien}]{eisenberg2013identifiability}
Eisenberg MC, Robertson SL, Tien JH (2013) Identifiability and estimation of
  multiple transmission pathways in cholera and waterborne disease. Journal of
  Theoretical Biology 324:84--102, \doi{10.1016/j.jtbi.2012.12.021}

\bibitem[{Engbert et~al.(2021)Engbert, Rabe, Kliegl, and
  Reich}]{engbert2021sequential}
Engbert R, Rabe MM, Kliegl R, Reich S (2021) {Sequential Data Assimilation of
  the Stochastic SEIR Epidemic Model for Regional COVID-19 Dynamics}. Bulletin
  of mathematical biology 83(1):1, \doi{10.1007/s11538-020-00834-8}

\bibitem[{Evans et~al.(2002)Evans, Chapman, Chappell, and
  Godfrey}]{evans2002identifiability}
Evans ND, Chapman MJ, Chappell MJ, Godfrey KR (2002) Identifiability of
  uncontrolled nonlinear rational systems. Automatica 38(10):1799--1805,
  \doi{10.1016/S0005-1098(02)00094-8}

\bibitem[{Fisman et~al.(2014)Fisman, Khoo, and Tuite}]{fisman2014early}
Fisman D, Khoo E, Tuite A (2014) {Early Epidemic Dynamics of the West African
  2014 Ebola Outbreak: Estimates Derived with a Simple Two-Parameter Model}.
  PLoS Currents 6,
  \doi{10.1371/currents.outbreaks.89c0d3783f36958d96ebbae97348d571}

\bibitem[{Fraser et~al.(2009)Fraser, Donnelly, Cauchemez, Hanage, Van~Kerkhove,
  Hollingsworth, Griffin, Baggaley, Jenkins, Lyons et~al.}]{fraser2009pandemic}
Fraser C, Donnelly CA, Cauchemez S, Hanage WP, Van~Kerkhove MD, Hollingsworth
  TD, Griffin J, Baggaley RF, Jenkins HE, Lyons EJ, et~al. (2009) {Pandemic
  potential of a strain of influenza A (H1N1): early findings}. Science
  324(5934):1557--1561, \doi{10.1126/science.1176062}

\bibitem[{Gao et~al.(2011)Gao, Liu, Nieto, and Andrade}]{gao2011seasonality}
Gao S, Liu Y, Nieto JJ, Andrade H (2011) Seasonality and mixed vaccination
  strategy in an epidemic model with vertical transmission. Mathematics and
  Computers in Simulation 81(9):1855--1868, \doi{10.1016/j.matcom.2010.10.032}

\bibitem[{Hong et~al.(2019)Hong, Ovchinnikov, Pogudin, and Yap}]{hong2019sian}
Hong H, Ovchinnikov A, Pogudin G, Yap C (2019) {SIAN: software for structural
  identifiability analysis of ODE models}. Bioinformatics 35(16):2873--2874,
  \doi{10.1093/bioinformatics/bty1069}

\bibitem[{Jacquez and Greif(1985)}]{jacquez1985numerical}
Jacquez JA, Greif P (1985) Numerical parameter identifiability and
  estimability: Integrating identifiability, estimability, and optimal sampling
  design. Mathematical Biosciences 77(1-2):201--227,
  \doi{10.1016/0025-5564(85)90098-7}

\bibitem[{Jacquez and Perry(1990)}]{jacquez1990parameter}
Jacquez JA, Perry T (1990) Parameter estimation: local identifiability of
  parameters. American Journal of Physiology-Endocrinology and Metabolism
  258(4):E727--E736, \doi{10.1152/ajpendo.1990.258.4.E727}

\bibitem[{Kao and Eisenberg(2018)}]{kao2018practical}
Kao YH, Eisenberg MC (2018) Practical unidentifiability of a simple
  vector-borne disease model: Implications for parameter estimation and
  intervention assessment. Epidemics 25:89--100,
  \doi{10.1016/j.epidem.2018.05.010}

\bibitem[{Li and Jin(2005)}]{li2005global}
Li G, Jin Z (2005) {Global stability of a SEIR epidemic model with infectious
  force in latent, infected and immune period}. Chaos, Solitons \& Fractals
  25(5):1177--1184, \doi{10.1016/j.chaos.2004.11.062}

\bibitem[{Li et~al.(2001)Li, Smith, and Wang}]{li2001global}
Li MY, Smith HL, Wang L (2001) {Global Dynamics of an SEIR Epidemic Model with
  Vertical Transmission}. SIAM Journal on Applied Mathematics 62(1):58--69,
  \doi{10.1137/S0036139999359860}

\bibitem[{Ligon et~al.(2018)Ligon, Fr{\"o}hlich, Chi{\c{s}}, Banga,
  Balsa-Canto, and Hasenauer}]{ligon2018genssi}
Ligon TS, Fr{\"o}hlich F, Chi{\c{s}} OT, Banga JR, Balsa-Canto E, Hasenauer J
  (2018) {GenSSI 2.0: multi-experiment structural identifiability analysis of
  SBML models}. Bioinformatics 34(8):1421--1423,
  \doi{10.1093/bioinformatics/btx735}

\bibitem[{Lipsitch et~al.(2003)Lipsitch, Cohen, Cooper, Robins, Ma, James,
  Gopalakrishna, Chew, Tan, Samore et~al.}]{lipsitch2003transmission}
Lipsitch M, Cohen T, Cooper B, Robins JM, Ma S, James L, Gopalakrishna G, Chew
  SK, Tan CC, Samore MH, et~al. (2003) Transmission dynamics and control of
  severe acute respiratory syndrome. Science 300(5627):1966--1970,
  \doi{10.1126/science.1086616}

\bibitem[{Liu(2019)}]{liu2019bifurcation}
Liu J (2019) {Bifurcation analysis for a delayed SEIR epidemic model with
  saturated incidence and saturated treatment function}. Journal of Biological
  Dynamics 13(1):461--480, \doi{10.1080/17513758.2019.1631965}

\bibitem[{Lizarralde-Bejarano et~al.(2020)Lizarralde-Bejarano, Rojas-D{\'\i}az,
  Arboleda-S{\'a}nchez, and Puerta-Yepes}]{lizarralde2020sensitivity}
Lizarralde-Bejarano DP, Rojas-D{\'\i}az D, Arboleda-S{\'a}nchez S, Puerta-Yepes
  ME (2020) {Sensitivity, uncertainty and identifiability analyses to define a
  dengue transmission model with real data of an endemic municipality of
  Colombia}. PLoS One 15(3):e0229,668, \doi{10.1371/journal.pone.0229668}

\bibitem[{Ljung and Glad(1994)}]{ljung1994global}
Ljung L, Glad T (1994) {On Global Identifiability for Arbitrary Model
  Parametrizations}. Automatica 30(2):265--276,
  \doi{10.1016/0005-1098(94)90029-9}

\bibitem[{Martynenko and B{\"u}ck(2018)}]{martynenko2018intelligent}
Martynenko A, B{\"u}ck A (2018) Intelligent control in drying. CRC Press,
  \doi{10.1201/9780429443183}

\bibitem[{Meshkat et~al.(2014)Meshkat, Kuo, and
  DiStefano~III}]{meshkat2014finding}
Meshkat N, Kuo CE, DiStefano~III J (2014) {On Finding and Using Identifiable
  Parameter Combinations in Nonlinear Dynamic Systems Biology Models and
  COMBOS: A Novel Web Implementation}. PLoS One 9(10):e110,261,
  \doi{10.1371/journal.pone.0110261}

\bibitem[{Metropolis and Ulam(1949)}]{metropolis1949monte}
Metropolis N, Ulam S (1949) {The Monte Carlo Method}. Journal of the American
  Statistical Association 44(247):335--341, \doi{10.2307/2280232}

\bibitem[{Miao et~al.(2011)Miao, Xia, Perelson, and
  Wu}]{miao2011identifiability}
Miao H, Xia X, Perelson AS, Wu H (2011) {On identifiability of nonlinear ODE
  models and applications in viral dynamics}. SIAM Review 53(1):3--39,
  \doi{10.1137/090757009}

\bibitem[{Pohjanpalo(1978)}]{pohjanpalo1978system}
Pohjanpalo H (1978) System identifiability based on the power series expansion
  of the solution. Mathematical Biosciences 41(1-2):21--33,
  \doi{10.1016/0025-5564(78)90063-9}

\bibitem[{Raue et~al.(2009)Raue, Kreutz, Maiwald, Bachmann, Schilling,
  Klingm{\"u}ller, and Timmer}]{raue2009structural}
Raue A, Kreutz C, Maiwald T, Bachmann J, Schilling M, Klingm{\"u}ller U, Timmer
  J (2009) Structural and practical identifiability analysis of partially
  observed dynamical models by exploiting the profile likelihood.
  Bioinformatics 25(15):1923--1929, \doi{10.1093/bioinformatics/btp358}

\bibitem[{Raue et~al.(2015)Raue, Steiert, Schelker, Kreutz, Maiwald, Hass,
  Vanlier, T{\"o}nsing, Adlung, Engesser et~al.}]{raue2015data2dynamics}
Raue A, Steiert B, Schelker M, Kreutz C, Maiwald T, Hass H, Vanlier J,
  T{\"o}nsing C, Adlung L, Engesser R, et~al. (2015) Data2dynamics: a modeling
  environment tailored to parameter estimation in dynamical systems.
  Bioinformatics 31(21):3558--3560

\bibitem[{Rey~Barreiro and Villaverde(2023)}]{rey2023benchmarking}
Rey~Barreiro X, Villaverde AF (2023) Benchmarking tools for a priori
  identifiability analysis. Bioinformatics 39(2):btad065,
  \doi{10.1093/bioinformatics/btad065}

\bibitem[{Riley et~al.(2003)Riley, Donnelly, and Ferguson}]{riley2003robust}
Riley S, Donnelly CA, Ferguson NM (2003) Robust parameter estimation techniques
  for stochastic within-host macroparasite models. Journal of Theoretical
  Biology 225(4):419--430, \doi{10.1016/s0022-5193(03)00266-2}

\bibitem[{Rodriguez-Fernandez et~al.(2006{\natexlab{a}})Rodriguez-Fernandez,
  Egea, and Banga}]{rodriguez2006novel}
Rodriguez-Fernandez M, Egea JA, Banga JR (2006{\natexlab{a}}) Novel
  metaheuristic for parameter estimation in nonlinear dynamic biological
  systems. BMC Bioinformatics 7(1):1--18, \doi{10.1186/1471-2105-7-483}

\bibitem[{Rodriguez-Fernandez et~al.(2006{\natexlab{b}})Rodriguez-Fernandez,
  Mendes, and Banga}]{rodriguez2006hybrid}
Rodriguez-Fernandez M, Mendes P, Banga JR (2006{\natexlab{b}}) A hybrid
  approach for efficient and robust parameter estimation in biochemical
  pathways. Biosystems 83(2-3):248--265, \doi{10.1016/j.biosystems.2005.06.016}

\bibitem[{Roosa and Chowell(2019)}]{roosa2019assessing}
Roosa K, Chowell G (2019) Assessing parameter identifiability in compartmental
  dynamic models using a computational approach: application to infectious
  disease transmission models. Theoretical Biology and Medical Modelling
  16(1):1--15

\bibitem[{Ruktanonchai et~al.(2020)Ruktanonchai, Floyd, Lai, Ruktanonchai,
  Sadilek, Rente-Lourenco, Ben, Carioli, Gwinn, Steele
  et~al.}]{ruktanonchai2020assessing}
Ruktanonchai NW, Floyd J, Lai S, Ruktanonchai CW, Sadilek A, Rente-Lourenco P,
  Ben X, Carioli A, Gwinn J, Steele J, et~al. (2020) {Assessing the impact of
  coordinated COVID-19 exit strategies across Europe}. Science
  369(6510):1465--1470, \doi{10.1126/science.abc5096}

\bibitem[{Sedoglavic(2002)}]{sedoglavic2002probabilistic}
Sedoglavic A (2002) {A Probabilistic Algorithm to Test Local Algebraic
  Observability in Polynomial Time}. Journal of Symbolic Computation
  33(5):735--755, \doi{10.1006/jsco.2002.0532}

\bibitem[{Tuite et~al.(2011)Tuite, Tien, Eisenberg, Earn, Ma, and
  Fisman}]{tuite2011cholera}
Tuite AR, Tien J, Eisenberg M, Earn DJ, Ma J, Fisman DN (2011) {Cholera
  epidemic in Haiti, 2010: using a transmission model to explain spatial spread
  of disease and identify optimal control interventions}. Annals of Internal
  Medicine 154(9):593--601, \doi{10.7326/0003-4819-154-9-201105030-00334}

\bibitem[{Tuncer and Le(2018)}]{tuncer2018structural}
Tuncer N, Le TT (2018) Structural and practical identifiability analysis of
  outbreak models. Mathematical Biosciences 299:1--18,
  \doi{10.1016/j.mbs.2018.02.004}

\bibitem[{Vajda et~al.(1989)Vajda, Godfrey, and Rabitz}]{vajda1989similarity}
Vajda S, Godfrey KR, Rabitz H (1989) Similarity transformation approach to
  identifiability analysis of nonlinear compartmental models. Mathematical
  Biosciences 93(2):217--248, \doi{10.1016/0025-5564(89)90024-2}

\bibitem[{Venzon and Moolgavkar(1988)}]{venzon1988method}
Venzon D, Moolgavkar S (1988) {A Method for Computing Profile-Likelihood-Based
  Confidence Intervals}. Journal of the Royal Statistical Society: Series C
  (Applied Statistics) 37(1):87--94, \doi{10.2307/2347496}

\bibitem[{Villaverde et~al.(2016)Villaverde, Barreiro, and
  Papachristodoulou}]{villaverde2016structural}
Villaverde AF, Barreiro A, Papachristodoulou A (2016) {Structural
  Identifiability of Dynamic Systems Biology Models}. PLoS Computational
  Biology 12(10):e1005,153, \doi{10.1371/journal.pcbi.1005153}

\bibitem[{Walter and Lecourtier(1981)}]{walter1981unidentifiable}
Walter E, Lecourtier Y (1981) Unidentifiable compartmental models: what to do?
  Mathematical Biosciences 56(1-2):1--25, \doi{10.1016/0025-5564(81)90025-0}

\bibitem[{Walter and Pronzato(1997)}]{walter1997identification}
Walter E, Pronzato L (1997) Identification of parametric models: from
  experimental data. Springer Verlag

\bibitem[{Walter et~al.(2004)Walter, Braems, Jaulin, and
  Kieffer}]{walter2004guaranteed}
Walter E, Braems I, Jaulin L, Kieffer M (2004) Guaranteed numerical computation
  as an alternative to computer algebra for testing models for identifiability.
  In: Numerical Software with Result Verification, Springer, pp 124--131,
  \doi{10.1007/b96498}

\bibitem[{Wieland et~al.(2021)Wieland, Hauber, Rosenblatt, T{\"o}nsing, and
  Timmer}]{wieland2021structural}
Wieland FG, Hauber AL, Rosenblatt M, T{\"o}nsing C, Timmer J (2021) On
  structural and practical identifiability. Current Opinion in Systems Biology
  25:60--69

\bibitem[{Wu et~al.(2008)Wu, Zhu, Miao, and Perelson}]{wu2008parameter}
Wu H, Zhu H, Miao H, Perelson AS (2008) Parameter identifiability and
  estimation of hiv/aids dynamic models. Bulletin of Mathematical Biology
  70(3):785--799, \doi{10.1007/s11538-007-9279-9}

\bibitem[{Xia and Moog(2003)}]{xia2003identifiability}
Xia X, Moog CH (2003) {Identifiability of Nonlinear Systems with Application to
  HIV/AIDS Models}. IEEE Transactions on Automatic Control 48(2):330--336,
  \doi{10.1109/TAC.2002.808494}

\bibitem[{Yan and Liu(2006)}]{yan2006seir}
Yan P, Liu S (2006) Seir epidemic model with delay. The ANZIAM Journal
  48(1):119--134, \doi{10.1017/S144618110000345X}

\bibitem[{Yates et~al.(2009)Yates, Evans, and Chappell}]{yates2009structural}
Yates JW, Evans ND, Chappell MJ (2009) Structural identifiability analysis via
  symmetries of differential equations. Automatica 45(11):2585--2591,
  \doi{10.1016/j.automatica.2009.07.009}

\end{thebibliography}

\section{Supplemental Figures}
\begin{figure}[H]
    \centering
    \includegraphics[width=10.35cm]{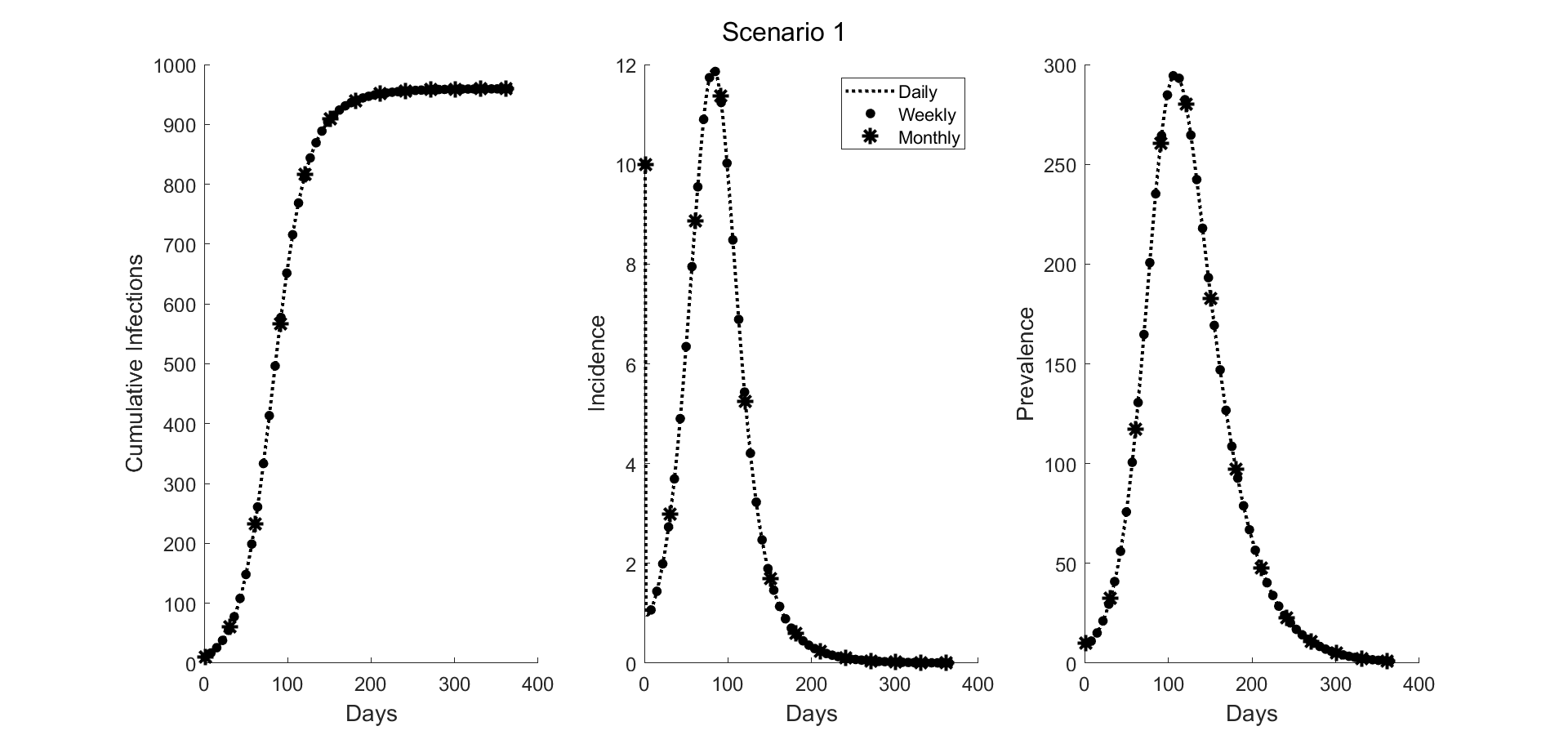} \\
      \includegraphics[width=10.35cm]{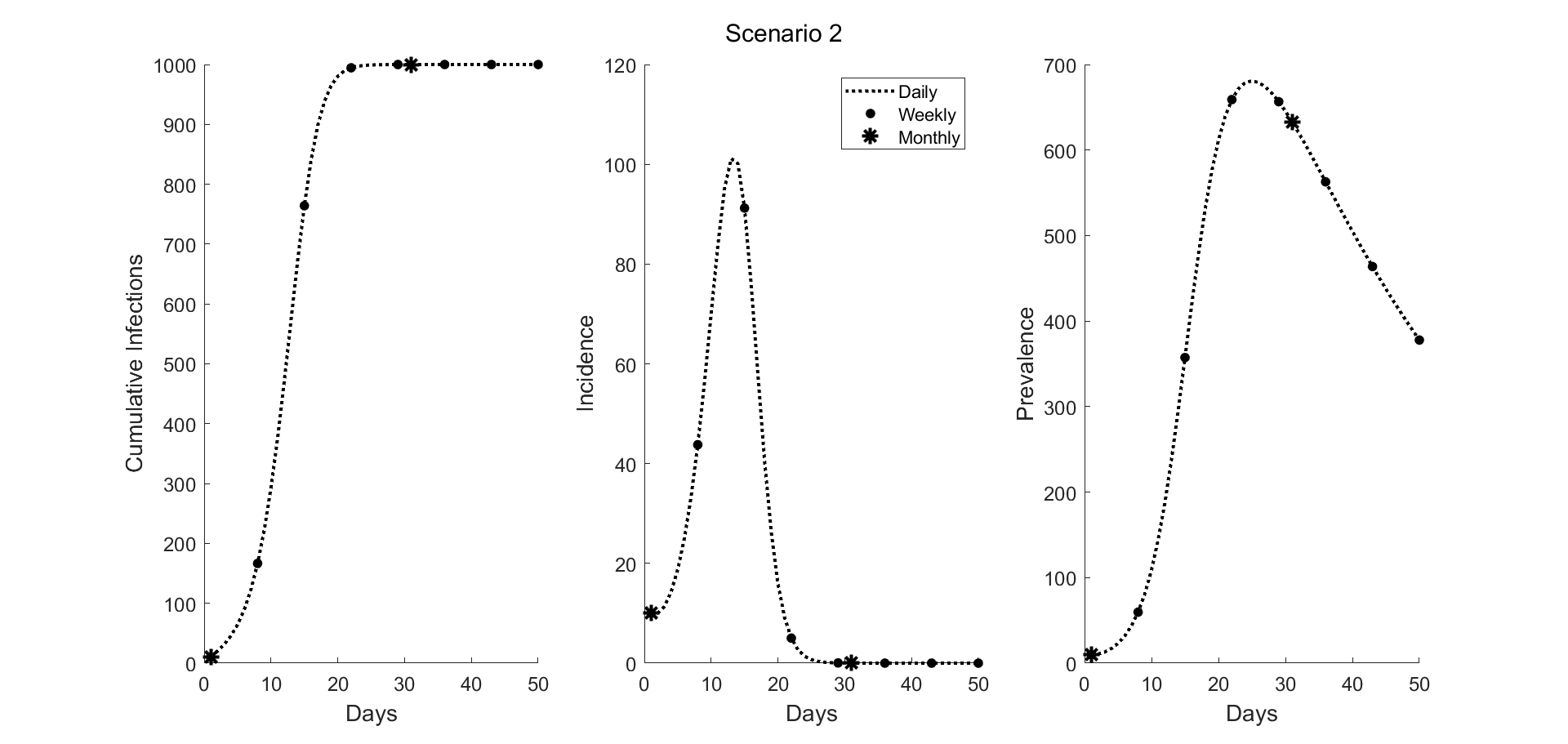} \\
        \includegraphics[width=10.35cm]{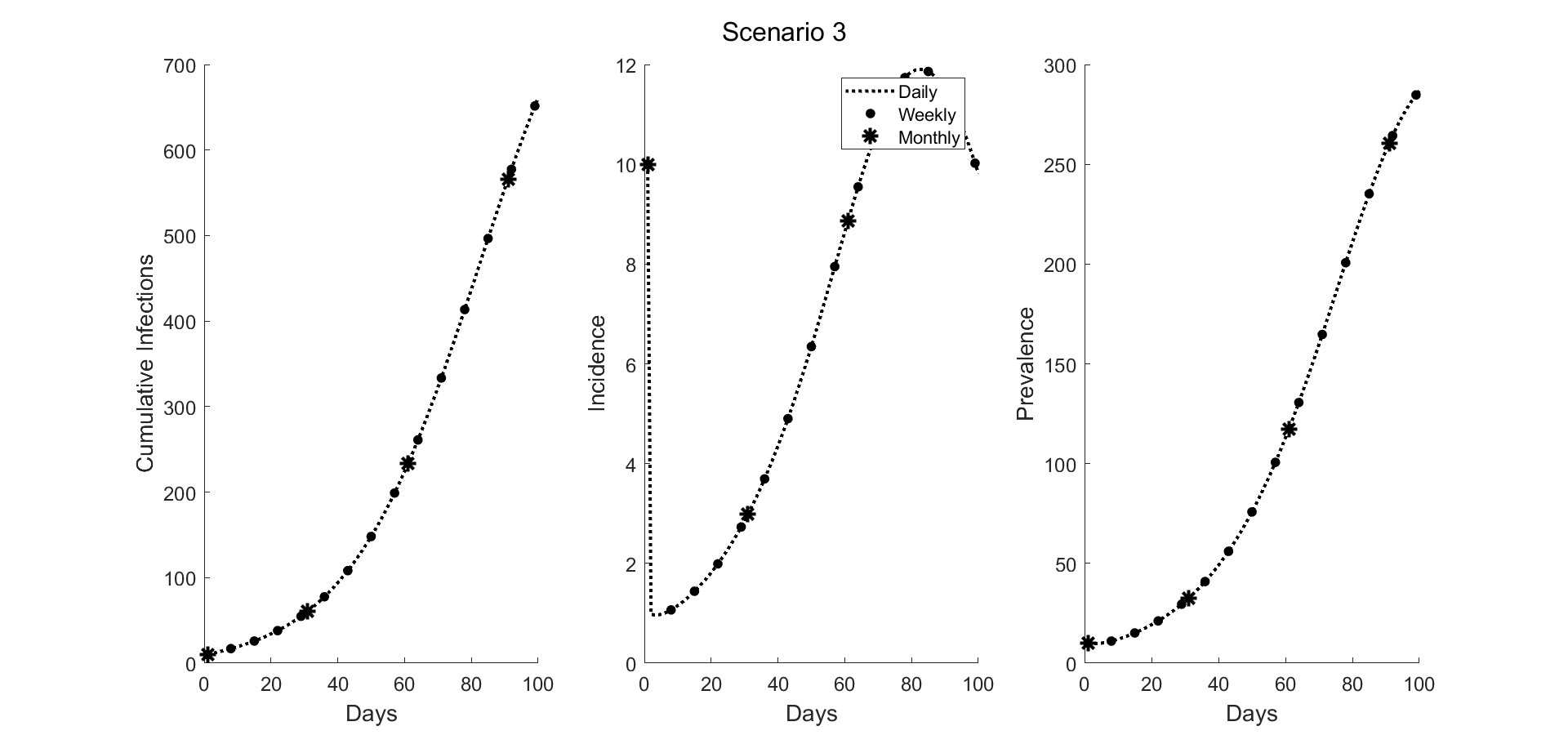} \\
          \includegraphics[width=10.35cm]{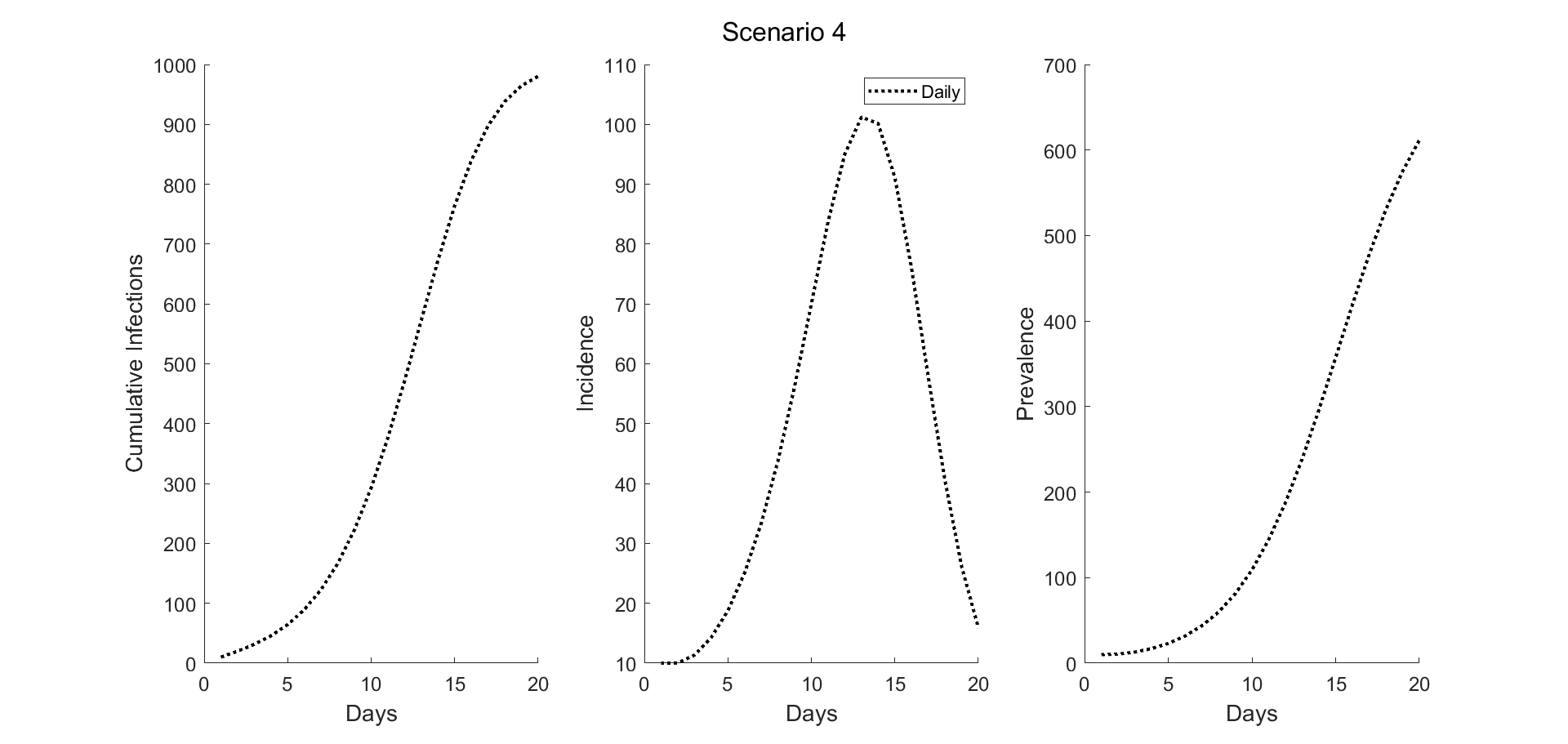} \\
    \caption{Plots for each scenario. In Scenario 1, the peak occurs on day 109 with the time span of 365 days.  In Scenario 2, the peak occurs on day 25 with the time span of 50 days. In Scenario 3, the peak still occurs on day 109, but the time span is reduced to 100 days. Similarly in Scenario 4, the peak is at day 25 with a time span of 20 days.  }
    \label{fig:scenarios}
\end{figure}

\begin{figure}[H]
    \centering
    \includegraphics[scale=.35]{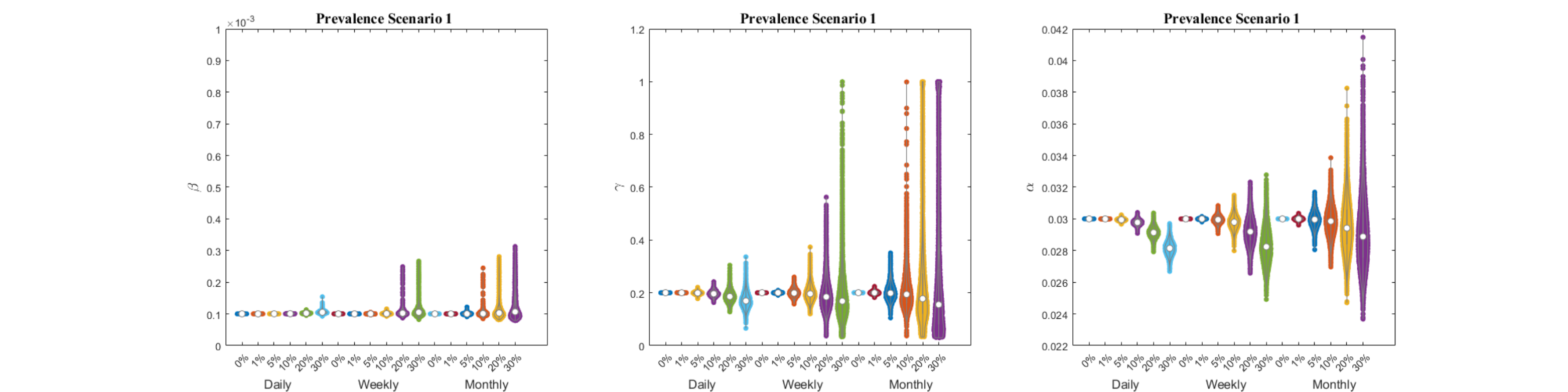} \\
      \includegraphics[scale=.35]{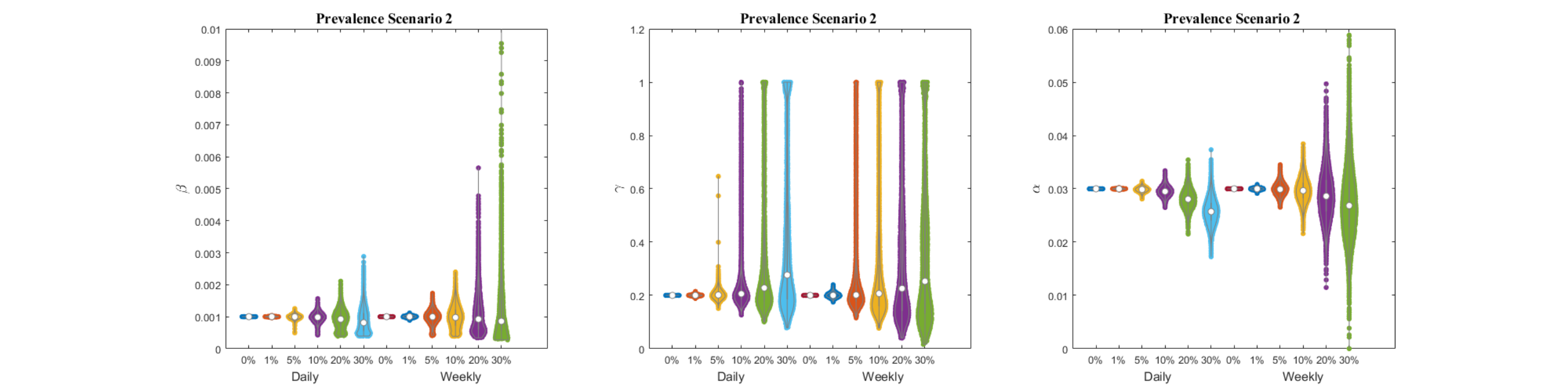} \\
        \includegraphics[scale=.35]{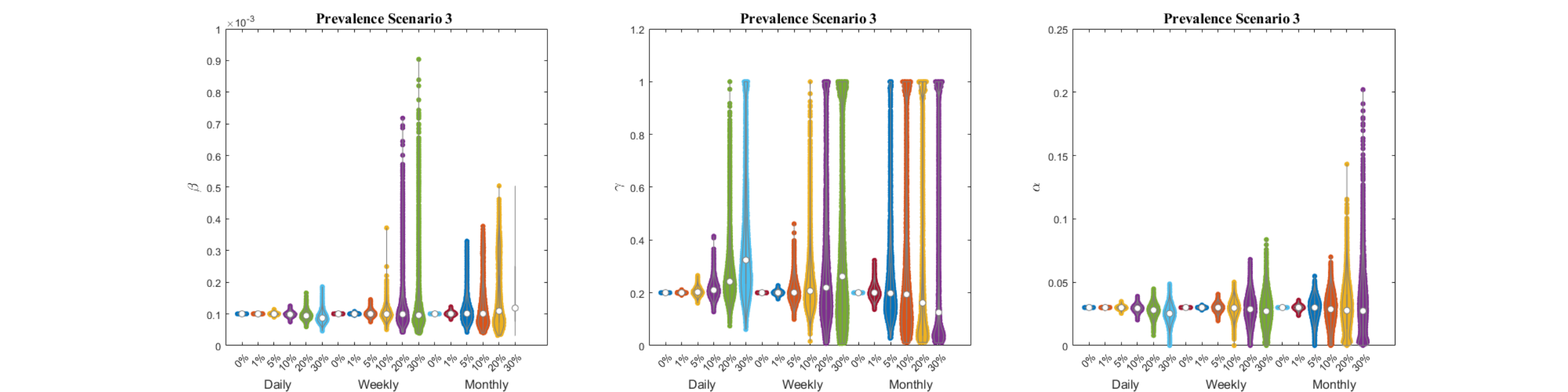} \\
          \includegraphics[scale=.35]{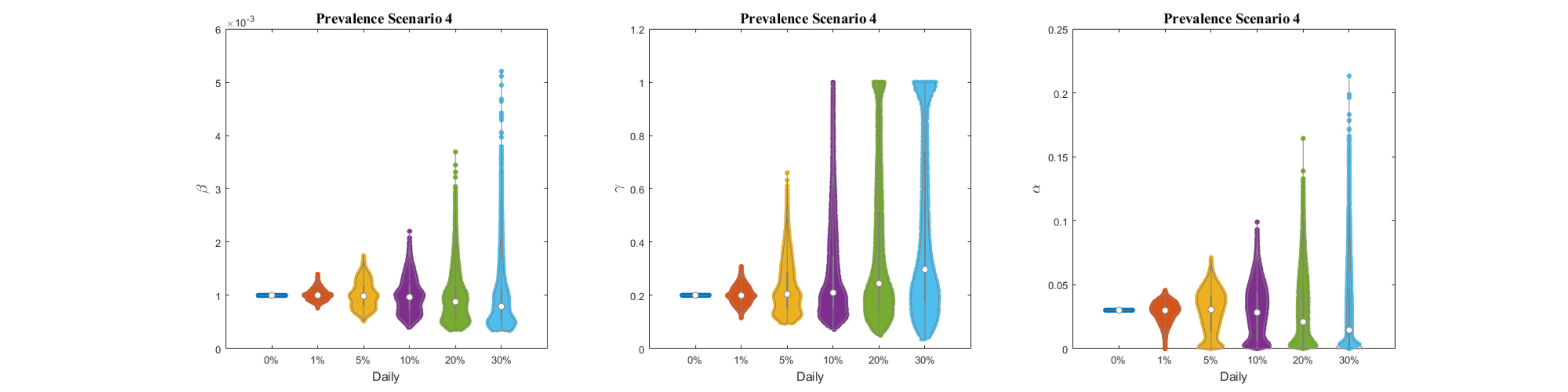} \\
    \caption{Violin Plots for $\beta$, $\gamma$, and $\alpha$ using Prevalence data for all four scenarios. These distributions are generated from MC algorithm using 10,000 iterations.}
    \label{fig:ViolinScenariosPrev}
\end{figure}

\newpage

The below figures plot normalized fval and normalized ARE for a given iteration versus other metrics of instance. Normalized ARE for a given scenario was obtained by dividing by the given ARE value by the noise level for the scenario. Thus normalized ARE values below 1 are considered identifiable while values greater than 1 are unidentifiable. The normalized fval was calculated as

\begin{align*}
   \frac{fval_{\text{True Params}}-fval_{\textbf{Est Params}}}{fval_{\text{True Params}}}.
\end{align*}
With this in mind positive values indicate the estimated parameter values fit the data better than the true parameter values while negative values indicate the true parameters provided a lower objective function value.

\begin{figure}
    \centering
    \includegraphics[scale=.425]{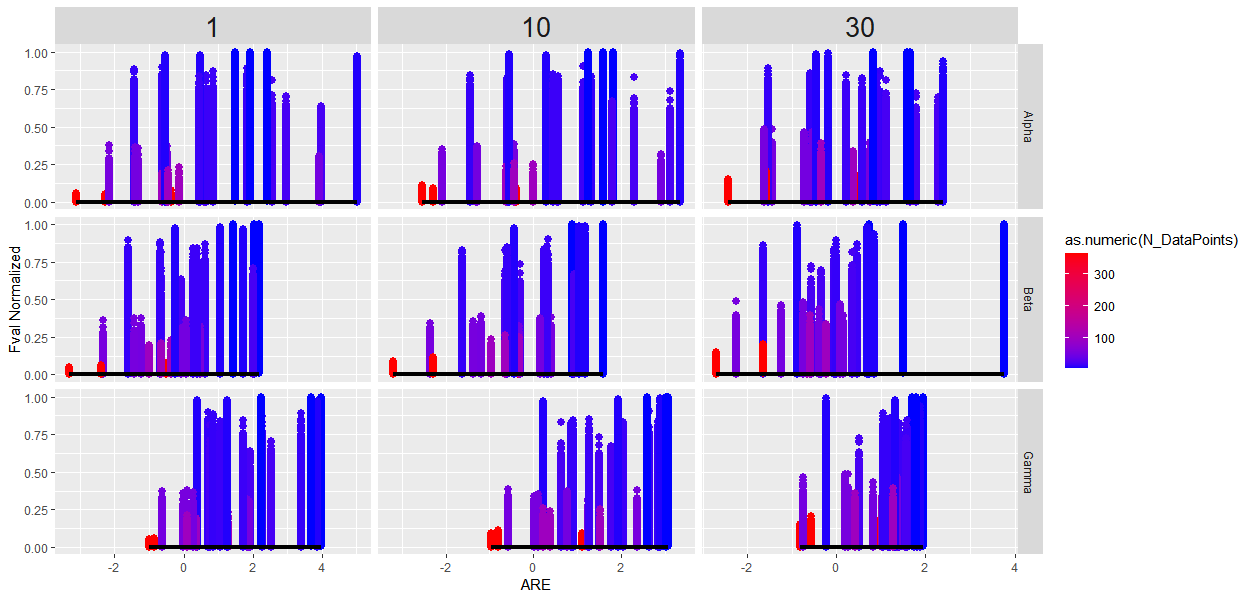}\\
 \caption{ARE on log scale versus normalized fval by noise level and data type. }
    \label{fig:ARE_vs_Fval}
\end{figure}

Figure \ref{fig:ARE_vs_Fval} shows ARE values plotted on a log scale against our normalized objective function value (fval). The data is further stratified by parameter type and noise level. The normalized objective function values tend to be above 0 indicating that the estimated parameters fit the noisy data better than the true parameter values. Scenarios with the lowest ARE and fval values with the smallest variation tend to correspond to scenarios with larger number of data points. This indicates that the optimization procedure struggles to identify the true parameters when noise is added to a small number of data points. Increasing ARE values tend to be associated with scenarios that have fewer data points and fvals that take on a larger range of values.

\begin{figure}[H]
    \centering
    \includegraphics[scale=.4]{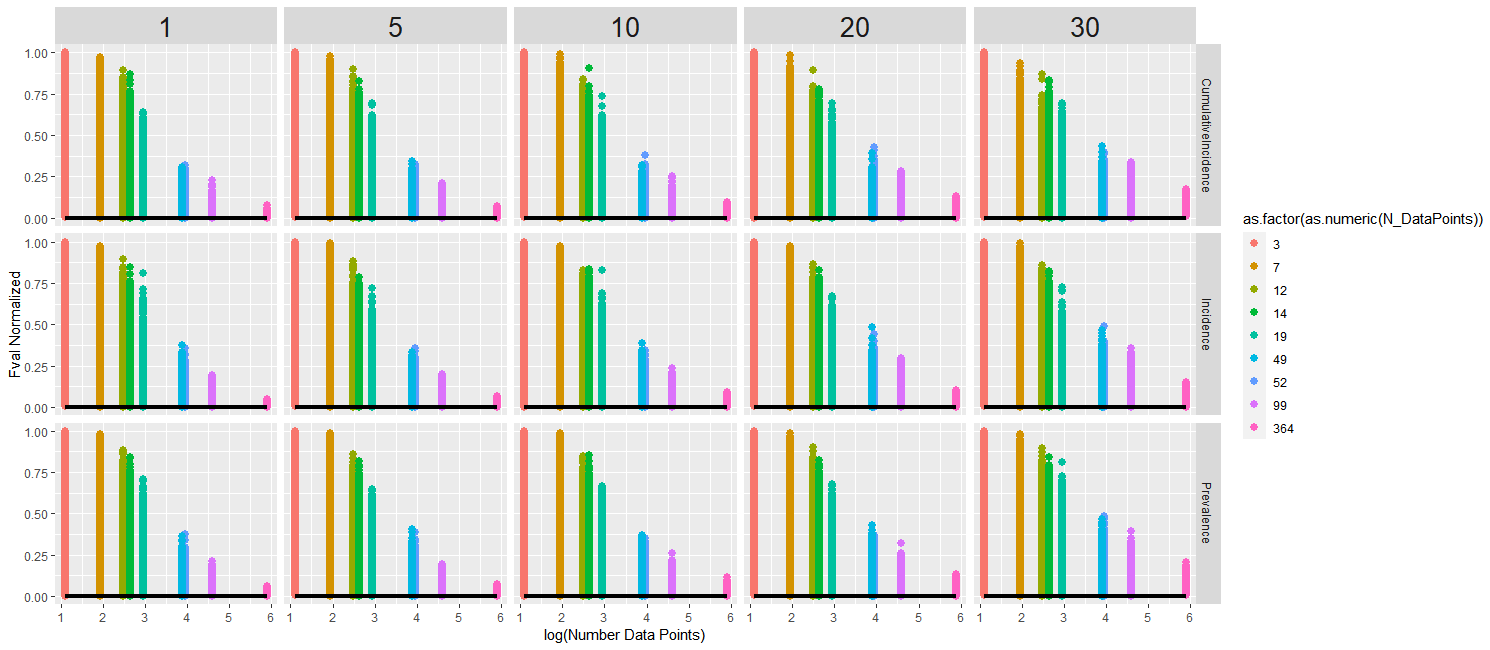}\\
    \caption{Number of data points on log scale versus fval normalized by noise level and data type. }
    \label{fig:NData_vs_fval}
\end{figure}

We notice similar dynamics in Figure \ref{fig:NData_vs_fval}, where the number of data point on a log scale is plotted against the normalized fval for different noise levels and data types. The normalized fval tending to be above 0 indicates that the estimated parameter values provide a better fit relative to true parameter values. As the number of data points decreases the fval shows a wider range, again demonstrating how adding noise to small number of data points complicates the parameter estimation process.

\begin{figure}[H]
    \centering
    \includegraphics[scale=.45]{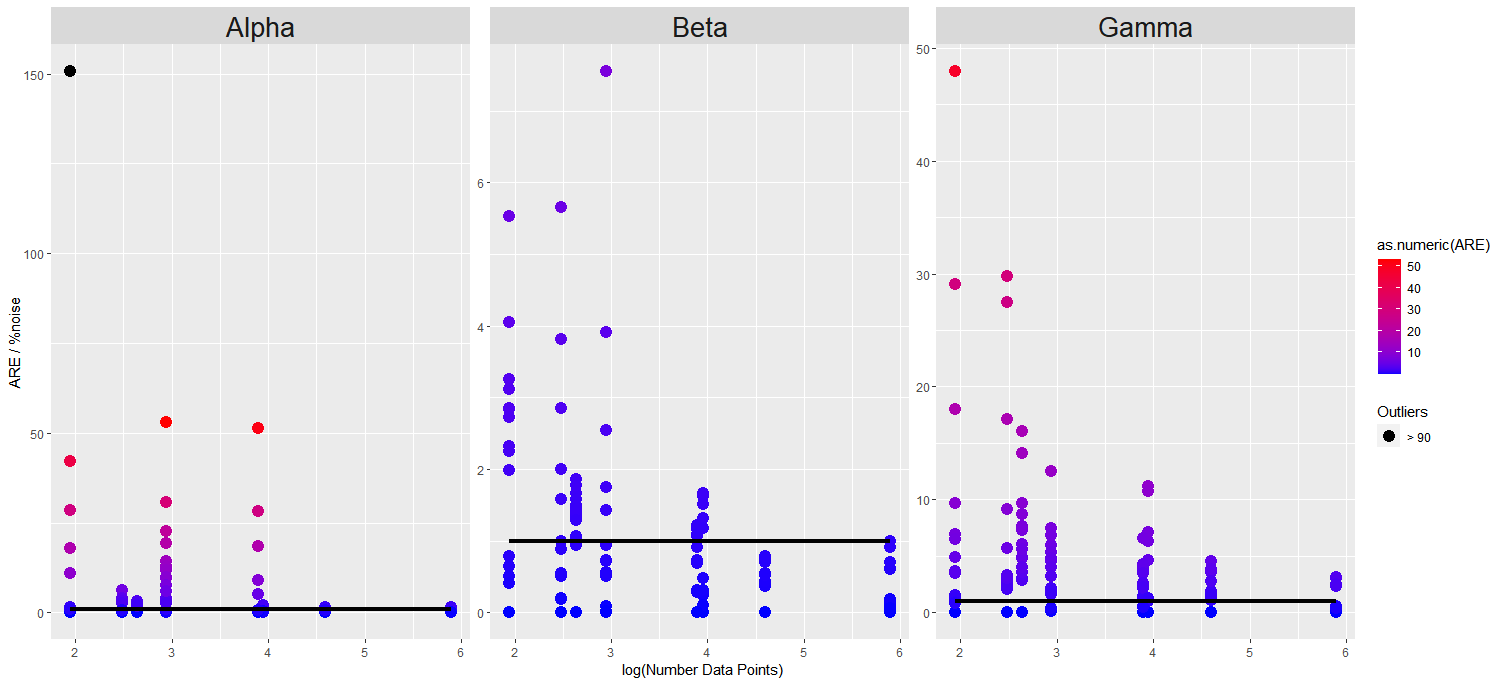}\\
    \caption{Number of data points on log scale versus normalized ARE by parameter. }
    \label{fig:NData_vs_ARE_param}
\end{figure}

\begin{figure}[H]
    \centering
    \includegraphics[scale=.45]{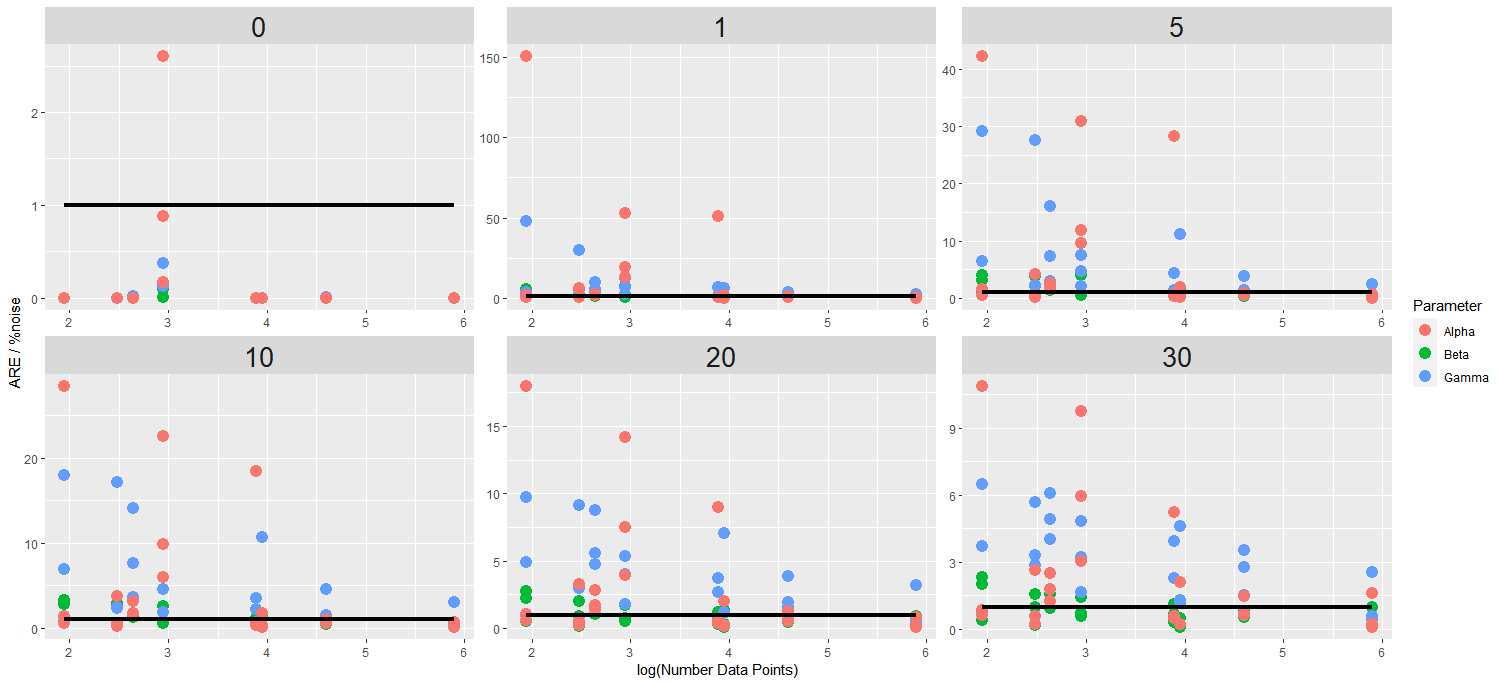}\\
    \caption{Number of data points on log scale versus normalized ARE by noise level.}
    \label{fig:NData_vs_ARE_noise}
\end{figure}

Figures \ref{fig:NData_vs_ARE_param} and \ref{fig:NData_vs_ARE_noise} plot the number of data points in a given scenario on the log scale versus ARE as broken down by parameter and noise level. ARE values below 1 are considered identifiable while values greater than 1 are unidentifiable. Beta clearly produced the lowest ARE values across all noise levels and data types. We see that as long as there are enough data points, beta is the most likely to be identifiable.

\begin{figure}[H]
   \centering
     \begin{subfigure}[b]{0.7\textwidth}
         \includegraphics[width=\textwidth]{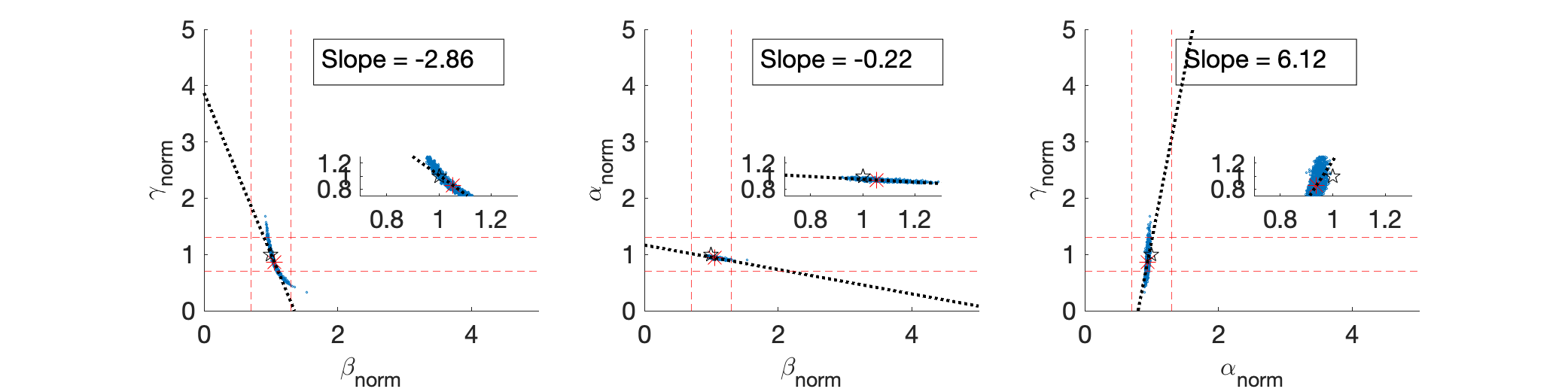}
         \caption{Prevalence, Scenario 1, Daily}
         \label{subfig:PrevS1D}
     \end{subfigure}

       \begin{subfigure}[b]{0.7\textwidth}
         \includegraphics[width=\textwidth]{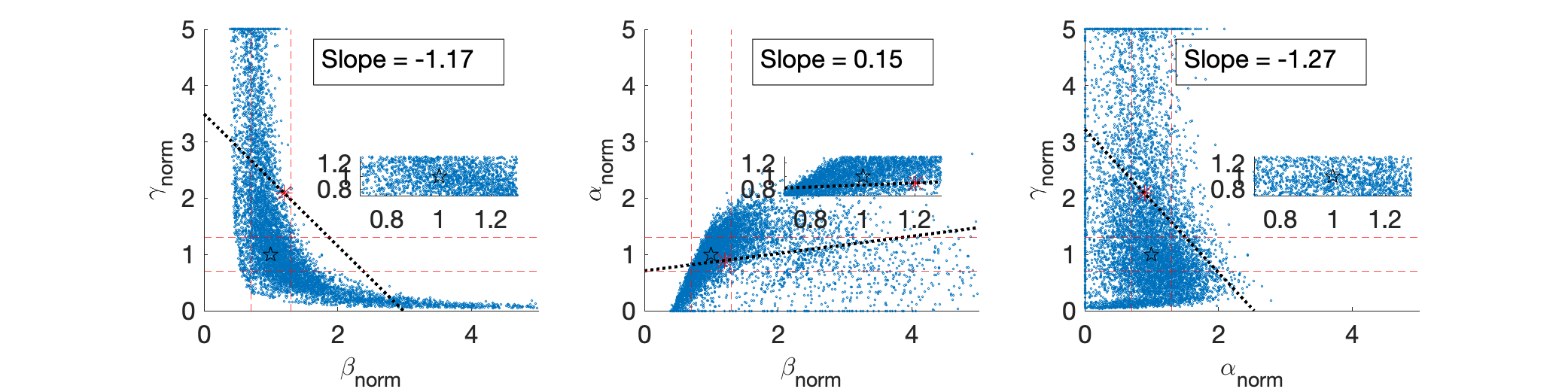}
         \caption{Prevalence, Scenario 3, Weekly}
         \label{subfig:PrevS3W}
     \end{subfigure}

     \begin{subfigure}[b]{0.7\textwidth}
      \includegraphics[width=\textwidth]{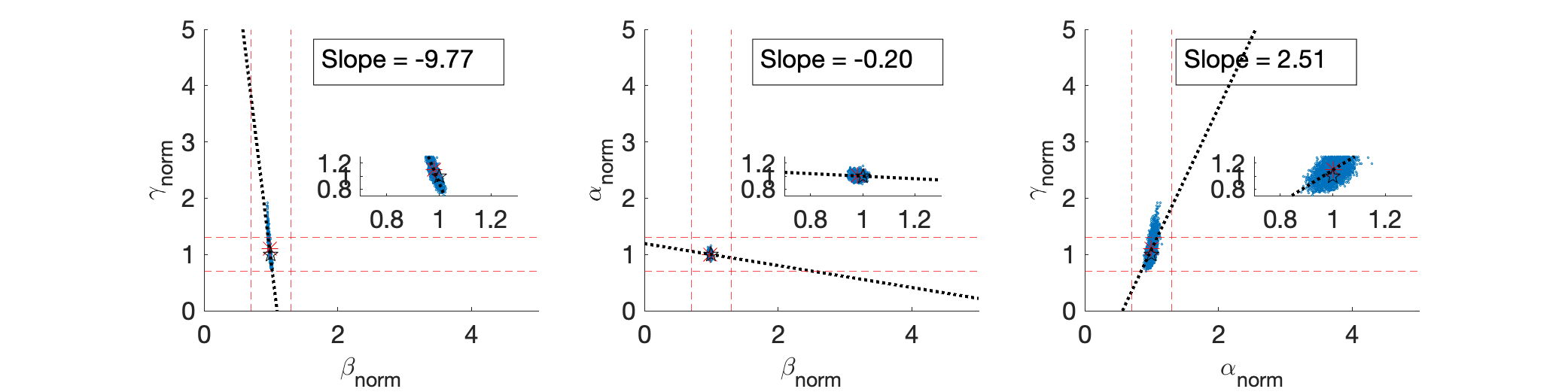}
         \caption{Incidence, Scenario 1, Daily}
         \label{subfig:IncS1D}
     \end{subfigure}

     \begin{subfigure}[b]{0.7\textwidth}
         \includegraphics[width=\textwidth]{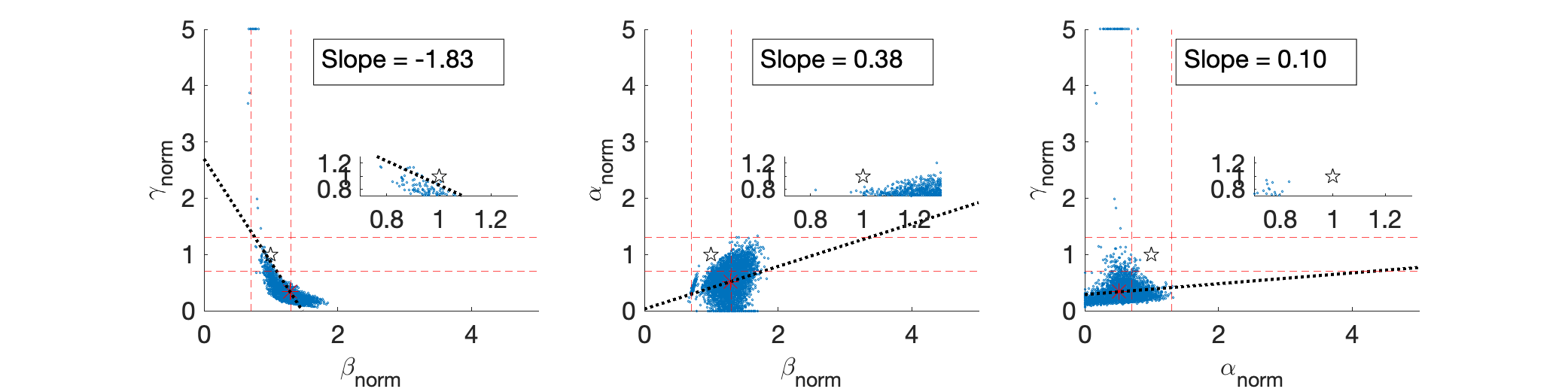}
         \caption{Cumulative Incidence, Scenario 1, Daily}
         \label{subfig:CumS1D}
     \end{subfigure}

     \begin{subfigure}[b]{0.7\textwidth}
         \includegraphics[width=\textwidth]{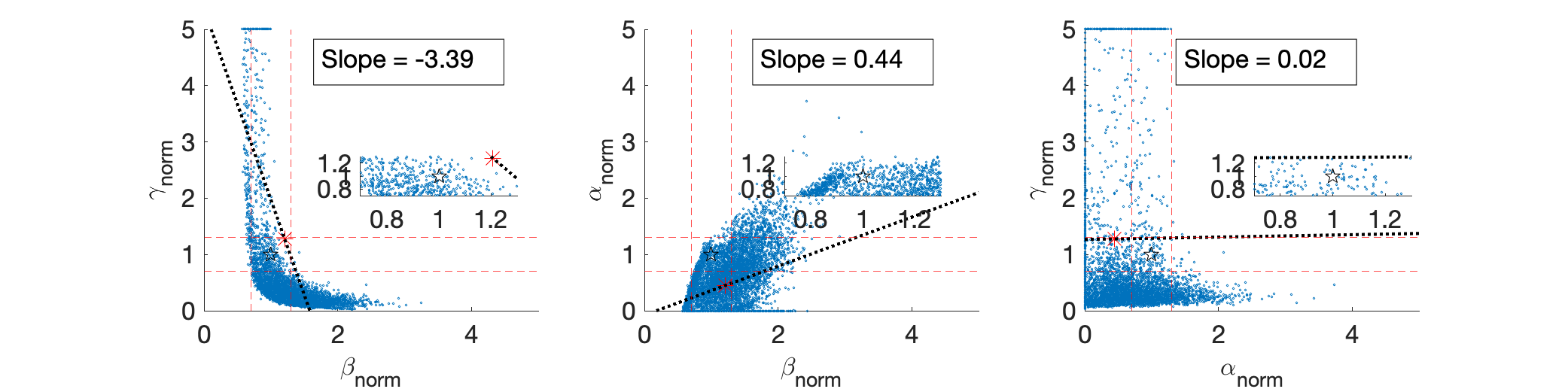}
         \caption{Cumulative Incidence, Scenario 1, Weekly}
         \label{subfig:CumS1W}
     \end{subfigure}
 \caption{Scatter plot of pairs of MC parameter estimates at 30\% noise level. Note that parameter estimates are normalized by the true value for comparison across select scenarios.}
        \label{fig:ScatterPlotMCParameterPairs}
\end{figure}

\newpage
\section{Supplemental Tables - Monte Carlo Results from \texttt{fminsearchbnd}}


\begin{table}[htb]
\centering
\begin{tabular}{|c|c c c|c c c|c c c|}
\hline
\bfseries Error level  & \multicolumn{3}{|c|}{\bfseries Daily } & \multicolumn{3}{|c|}{\bfseries Weekly}& \multicolumn{3}{|c|}{\bfseries Monthly }\\
\hline
\hline
$\bm{\sigma}$ & $\bm{\beta}$ &$\bm{\gamma}$ &$\bm{\alpha}$ & $\bm{\beta}$ &$\bm{\gamma}$ &$\bm{\alpha}$ & $\bm{\beta}$ &$\bm{\gamma}$ &$\bm{\alpha}$\\
\hline
0 \%& 0.00 & 0.00 & 0.00 & 0.00 & 0.00 & 0.00 & 0.00 & 0.00 & 0.00 \\
1 \%& 0.09 & 0.42 & 0.04 & 0.24 & 1.10 & 0.11 & 0.51 & 2.37 & 0.24 \\
5 \% & 0.47 & 2.16 & 0.26 & 1.21 & 5.55 & 0.59 & 2.50 & 11.64 & 1.18 \\
10 \% & 1.03 & 4.45 & 0.79 & 2.50 & 11.20 & 1.25 & 5.52 & 24.69 & 2.36 \\
20\% & 2.69 & 9.96 & 2.88 & 6.27 & 23.12 & 3.14 & 17.76 & 59.51 & 4.58 \\
30 \%& 5.77 & 17.15 & 6.19 & 14.28 & 36.38 & 5.92 & 29.70 & 85.60 & 6.51 \\
\hline
\bfseries Identifiable  &\cellcolor[HTML]{8EA9DB} Yes & \cellcolor[HTML]{8EA9DB}Yes &\cellcolor[HTML]{8EA9DB} Yes & \cellcolor[HTML]{8EA9DB}Yes & No & \cellcolor[HTML]{8EA9DB}Yes & \cellcolor[HTML]{8EA9DB}Yes & No & \cellcolor[HTML]{8EA9DB}Yes\\
\hline
\end{tabular}
\vspace{-0.20cm}
\caption{Prevalence Scenario 1}

\label{Prevalence Scenario 1 FMSB}
\end{table}
\vspace{-0.30cm}
\begin{table}[H]
\centering
\caption{Prevalence Scenario 2}
\begin{tabular}{|c|c c c|c c c|}
\hline
\bfseries Error level & \multicolumn{3}{|c|}{\bfseries Daily } & \multicolumn{3}{|c|}{\bfseries Weekly}\\
\hline
\hline
$\bm{\sigma}$ & $\bm{\beta}$ &$\bm{\gamma}$ &$\bm{\alpha}$ & $\bm{\beta}$ &$\bm{\gamma}$ &$\bm{\alpha}$\\
\hline
0 \%& 0.00 & 0.00 & 0.00 &  0.00 & 0.00 & 0.00 \\
1 \%& 1.06 & 1.32 & 0.24 &  2.85 & 3.53 & 0.58 \\
5 \%& 5.43 & 7.03 & 1.27 &  15.60 & 32.60 & 2.89 \\
10\% & 11.69 & 21.42 & 2.76 &  28.37 & 69.06 & 5.80 \\
20 \%& 24.40 & 74.11 & 7.25 &  45.12 & 98.03 & 12.26 \\
30 \%& 32.74 & 117.37 & 14.49 &  59.74 & 110.75 & 19.18 \\
\hline
\bfseries Identifiable  & No & No & \cellcolor[HTML]{8EA9DB}Yes & No & No & \cellcolor[HTML]{8EA9DB}Yes \\
\hline
\end{tabular}

\vspace{-0.20cm}
\label{Prevalence Scenario 2 FMSB}
\end{table}

\vspace{-0.30cm}

\begin{table}[H]
\centering

\begin{tabular}{|c|c c c|c c c|c c c|}
\hline
\bfseries Error level  & \multicolumn{3}{|c|}{\bfseries Daily } & \multicolumn{3}{|c|}{\bfseries Weekly }& \multicolumn{3}{|c|}{\bfseries Monthly}\\
\hline
\hline
$\bm{\sigma}$ & $\bm{\beta}$ &$\bm{\gamma}$ &$\bm{\alpha}$ & $\bm{\beta}$ &$\bm{\gamma}$ &$\bm{\alpha}$ & $\bm{\beta}$ &$\bm{\gamma}$ &$\bm{\alpha}$\\
\hline
0\% & 0.00 & 0.00 & 0.00 & 0.00 & 0.00 & 0.00 & 0.00 & 0.00 & 0.00 \\
1\% & 0.53 & 1.11 & 0.65 & 1.32 & 2.89 & 1.56 & 4.25 & 9.36 & 4.42 \\
5\%& 2.63 & 5.75 & 3.24 & 6.69 & 15.01 & 7.86 & 23.00 & 70.17 & 21.24 \\
10\% & 5.21 & 12.55 & 6.48 & 13.81 & 35.67 & 15.69 & 49.20 & 133.19 & 34.41 \\
20\% & 10.72 & 38.33 & 13.45 & 33.24 & 95.75 & 28.93 & 76.36 & 168.39 & 52.20 \\
30\% & 16.44 & 106.42 & 21.23 & 47.27 & 147.87 & 37.13 & 1287.14 & 178.02 & 68.08 \\
\hline
\bfseries Identifiable  & \cellcolor[HTML]{8EA9DB}Yes & No & \cellcolor[HTML]{8EA9DB}Yes & No & No & No  &  No & No & No\\
\hline
\end{tabular}
\vspace{-0.20cm}
\caption{Prevalence Scenario 3}
\label{Prevalence Scenario 3 FMSB}
\end{table}

\vspace{-0.30cm}

\begin{table}[H]
\centering
\begin{tabular}{|c|c c c|}
\hline
\bfseries Error level  & \multicolumn{3}{|c|}{\bfseries Daily }\\
\hline
\hline
$\bm{\sigma}$ & $\bm{\beta}$ &$\bm{\gamma}$ &$\bm{\alpha}$\\
\hline
0\% & 0.09 & 0.14 & 0.17 \\
1\% & 7.55 & 12.52 & 19.24 \\
5\% & 19.57 & 37.17 & 48.05 \\
10\% & 25.50 & 59.28 & 59.90 \\
20\% & 34.91 & 106.81 & 78.40 \\
30\% & 43.04 & 144.20 & 91.45 \\
\hline
\bfseries Identifiable & No & No & No \\
\hline
\end{tabular}

\caption{Prevalence Scenario 4}
\label{Prevalence Scenario 4 FMSB}
\end{table}

\begin{table}[H]
\centering
\begin{tabular}{|p{1cm}|c c c|c c c|c c c|}
\hline
\bfseries Error level & \multicolumn{3}{|c|}{\bfseries Daily} & \multicolumn{3}{|c|}{\bfseries Weekly}& \multicolumn{3}{|c|}{\bfseries Monthly }\\
\hline
\hline
$\bm{\sigma}$ & $\bm{\beta}$ &$\bm{\gamma}$ &$\bm{\alpha}$ & $\bm{\beta}$ &$\bm{\gamma}$ &$\bm{\alpha}$ & $\bm{\beta}$ &$\bm{\gamma}$ &$\bm{\alpha}$\\
\hline
0\% & 0.00 & 0.00 & 0.00 &0.00 &0.00 & 0.00 & 0.00 & 0.00 & 0.00 \\
1\% & 0.04 & 0.37 & 0.10 & 0.10 & 0.99 & 0.27 & 0.20 & 2.03 & 0.53 \\
5\% & 0.19 & 1.87 & 0.52 & 0.48 & 4.92 & 1.36 & 0.98 & 10.38 & 2.70 \\
10\% & 0.41 & 3.81 & 1.03 & 0.96 & 10.07 & 2.68 & 1.97 & 23.36 & 5.47 \\
20\% & 1.03 & 8.07 & 1.97 & 1.99 & 22.32 & 5.13 & 3.75 & 61.51 & 10.80 \\
30\% & 2.03 & 13.36 & 2.65 & 3.13 & 38.75 & 7.08 & 5.73 & 99.28 & 16.78 \\
\hline
\bfseries Identifiable  & \cellcolor[HTML]{8EA9DB}Yes & \cellcolor[HTML]{8EA9DB}Yes & \cellcolor[HTML]{8EA9DB}Yes  & \cellcolor[HTML]{8EA9DB}Yes & No &\cellcolor[HTML]{8EA9DB} Yes  & \cellcolor[HTML]{8EA9DB}Yes & No & \cellcolor[HTML]{8EA9DB}Yes \\
\hline
\end{tabular}
\vspace{-0.20cm}

\caption{Incidence Scenario 1}
\label{Incidence Scenario 1 FMSB}
\end{table}

\vspace{-0.30cm}
\begin{table}[H]
\centering
\begin{tabular}{|c|c c c|c c c|}
\hline
\bfseries Error level  & \multicolumn{3}{|c|}{\bfseries Daily } & \multicolumn{3}{|c|}{\bfseries Weekly }\\
\hline
\hline
$\bm{\sigma}$ & $\bm{\beta}$ &$\bm{\gamma}$ &$\bm{\alpha}$ & $\bm{\beta}$ &$\bm{\gamma}$ &$\bm{\alpha}$ \\
\hline
  0\% & 0.00 & 0.00 & 0.00 & 0.00 & 0.00 & 0.00 \\
        1\% & 0.29 & 0.54 & 0.61 & 0.78 & 1.49 & 1.60 \\
        5\% & 1.46 & 2.73 & 3.07 & 3.95 & 7.46 & 8.15 \\
        10\% & 3.07 & 5.64 & 6.45 & 6.45 & 12.44 & 13.25 \\
        20\% & 5.64 & 10.23 & 11.80 & 10.01 & 19.21 & 20.48 \\
        30\% & 2.03 & 13.36 & 2.65 & 12.36 & 24.00 & 24.94 \\
\hline
\bfseries Identifiable  &  \cellcolor[HTML]{8EA9DB}Yes & \cellcolor[HTML]{8EA9DB}Yes & \cellcolor[HTML]{8EA9DB}Yes & \cellcolor[HTML]{8EA9DB}Yes & \cellcolor[HTML]{8EA9DB}Yes & \cellcolor[HTML]{8EA9DB}Yes \\
\hline
\end{tabular}
\vspace{-0.20cm}

\caption{Incidence Scenario 2 }
\label{ Incidence Scenario 2 FMSB}
\end{table}

\vspace{-0.30cm}

\begin{table}[H]
\centering
\begin{tabular}{|c|c c c|c c c|c c c|}
\hline
\bfseries Error level  & \multicolumn{3}{|c|}{\bfseries Daily } & \multicolumn{3}{|c|}{\bfseries Weekly }& \multicolumn{3}{|c|}{\bfseries Monthly }\\
\hline
\hline
$\bm{\sigma}$ & $\bm{\beta}$ &$\bm{\gamma}$ &$\bm{\alpha}$ & $\bm{\beta}$ &$\bm{\gamma}$ &$\bm{\alpha}$ & $\bm{\beta}$ &$\bm{\gamma}$ &$\bm{\alpha}$\\
\hline
0\% & 0.00 & 0.00 & 0.00 & 0.00 & 0.02 & 0.00 & 0.00 & 0.00 & 0.00 \\
1\% & 0.37 & 1.45 & 0.54 & 1.50 & 5.65 & 1.89 & 8.91 & 40.28 & 11.26 \\
5\% & 1.86 & 7.32 & 2.71 & 7.26 & 36.33 & 9.56 & 32.58 & 175.46 & 32.72 \\
10 \%& 3.82 & 14.65 & 5.39 & 12.90 & 76.38 & 17.62 & 48.46 & 205.18 & 49.26 \\
20\% & 8.73 & 30.68 & 11.01 & 21.30 & 111.74 & 33.87 & 75.24 & 193.70 & 90.65 \\
30\% & 15.51 & 45.53 & 19.16 &  28.22 & 121.14 & 53.24 & 133.93 & 164.62 & 156.34 \\
\hline
\bfseries Identifiable  &  \cellcolor[HTML]{8EA9DB}Yes & No &  \cellcolor[HTML]{8EA9DB}Yes &   \cellcolor[HTML]{8EA9DB}Yes & No & No & No & No & No \\
\hline
\end{tabular}
\vspace{-0.20cm}

\caption{Incidence Scenario 3 }
\label{Incidence Scenario 3 FMSB}
\end{table}
\vspace{-0.30cm}

\begin{table}[H]
\centering
\begin{tabular}{|c|c c c|}
\hline
\bfseries Error level  & \multicolumn{3}{|c|}{\bfseries Daily }\\
\hline
\hline
$\bm{\sigma}$ & $\bm{\beta}$ &$\bm{\gamma}$ &$\bm{\alpha}$\\
\hline
0\% & 2.26 & 3.93 & 5.46 \\
1\% & 7.46 & 13.18 & 18.50 \\
5\%& 19.16 & 38.24 & 45.39 \\
10\% & 25.85 & 60.22 & 57.45 \\
20\% & 36.36 & 103.55 & 76.31 \\
30\% & 45.01 & 133.93 & 90.46 \\
\hline
\bfseries Identifiable & No & No & No \\
\hline
\end{tabular}
\caption{Incidence  Scenario 4 }
\label{Incidence Scenario 4 FMSB}
\end{table}
\vspace{-0.30cm}

\begin{table}[H]
\centering
\begin{tabular}{|c|c c c|c c c|c c c|}
\hline
\bfseries Error level  & \multicolumn{3}{|c|}{\bfseries Daily } & \multicolumn{3}{|c|}{\bfseries Weekly }& \multicolumn{3}{|c|}{\bfseries Monthly }\\
\hline
\hline
$\bm{\sigma}$ & $\bm{\beta}$ &$\bm{\gamma}$ &$\bm{\alpha}$ & $\bm{\beta}$ &$\bm{\gamma}$ &$\bm{\alpha}$ & $\bm{\beta}$ &$\bm{\gamma}$ &$\bm{\alpha}$\\
\hline
0\% & 0.00 & 0.00 & 0.00 & 0.00 & 0.00 & 0.00 & 0.00 & 0.00 & 0.00 \\
1\% & 0.60 & 2.34 & 0.69 & 1.62 & 6.34 & 1.83 & 5.65 & 29.82 & 6.21 \\
5\%& 3.12 & 12.04 & 3.39 & 8.31 & 55.85 & 9.40 & 19.07 & 137.41 & 20.68 \\
10 \%& 7.08 & 30.59 & 6.80 & 15.07 & 107.18 & 17.29 & 28.51 & 170.83 & 37.45 \\
20 \%& 18.37 & 63.31 & 16.27 & 26.41 & 141.72 & 40.16 & 39.89 & 182.41 & 64.83 \\
30\% & 29.67 & 76.41 & 48.44 & 35.10 & 138.42 & 63.45 & 47.30 & 170.29 & 79.50 \\
\hline
\bfseries Identifiable  &  \cellcolor[HTML]{8EA9DB}Yes& No & No & No & No & No & No & No & No \\
\hline
\end{tabular}
\vspace{-0.20cm}

\caption{Cumulative Incidence Scenario 1}
\label{Cumulative Incidence Scenario 1 FMSB}
\end{table}
\vspace{-0.30cm}
\begin{table}[H]
\centering
\begin{tabular}{|c|c c c|c c c|}
\hline
\bfseries Error level  & \multicolumn{3}{|c|}{\bfseries Daily } & \multicolumn{3}{|c|}{\bfseries Weekly }\\
\hline
\hline
$\bm{\sigma}$ & $\bm{\beta}$ &$\bm{\gamma}$ &$\bm{\alpha}$ & $\bm{\beta}$ &$\bm{\gamma}$ &$\bm{\alpha}$ \\
\hline
  0\% & 0.00 & 0.00 & 0.00 & 0.00 & 0.00 & 0.00\\
        1\% & 0.92 & 6.59 & 51.18 & 5.53 & 47.91 & 150.65 \\
        5\% & 3.62 & 21.28 & 141.23 & 20.28 & 145.39 & 210.88\\
        10\% & 7.19 & 34.65 & 183.97 & 32.61 & 179.75 & 284.48 \\
        20\% & 14.00 & 52.83 & 179.59 & 54.39 & 193.97 & 359.43 \\
        30\% & 20.50 & 68.05 & 156.92 & 69.78 & 195.01 & 326.25 \\
\hline
\bfseries Identifiable  &  \cellcolor[HTML]{8EA9DB}Yes & No & No & No & No & No \\
\hline
\end{tabular}
\vspace{-0.20cm}

\caption{Cumulative Incidence Scenario 2 }
\label{Cumulative Incidence Scenario 2 FMSB}
\end{table}

\vspace{-0.30cm}
\begin{table}[H]
\centering
\begin{tabular}{|c|c c c|c c c|c c c|}
\hline
\bfseries Error level  & \multicolumn{3}{|c|}{\bfseries Daily } & \multicolumn{3}{|c|}{\bfseries Weekly }& \multicolumn{3}{|c|}{\bfseries Monthly }\\
\hline
\hline
$\bm{\sigma}$ & $\bm{\beta}$ &$\bm{\gamma}$ &$\bm{\alpha}$ & $\bm{\beta}$ &$\bm{\gamma}$ &$\bm{\alpha}$ & $\bm{\beta}$ &$\bm{\gamma}$ &$\bm{\alpha}$\\
\hline
0\% & 0.00 & 0.00 & 0.00 & 0.00 & 0.00 & 0.00 & 0.00 & 0.00 & 0.00 \\
1 \%& 0.70 & 3.62 & 0.89 &  1.86 & 9.72 & 2.33 &  7.60 & 54.17 & 6.95 \\
5 \%& 3.50 & 19.11 & 4.53 &  8.90 & 80.27 & 13.36 &  19.24 & 188.96 & 37.96 \\
10\% & 7.19 & 45.52 & 9.93 & 13.98 & 140.76 & 30.95 &  24.06 & 215.03 & 61.91 \\
20 \%& 14.53 & 77.25 & 25.55 & 20.70 & 174.66 & 55.82 &  39.16 & 216.27 & 95.20 \\
30 \%& 23.70 & 83.67 & 43.67 &  28.60 & 182.85 & 74.99 &  62.51 & 210.04 & 145.29 \\
\hline
\bfseries Identifiable  &  \cellcolor[HTML]{8EA9DB}Yes & No & No &  \cellcolor[HTML]{8EA9DB}Yes & No & No & No & No & No \\
\hline
\end{tabular}
\vspace{-0.20cm}

\caption{Cumulative Incidence Scenario 3 }
\label{Cumulative Incidence Scenario 3 FMSB}
\end{table}

\vspace{-0.30cm}

\begin{table}[H]
\centering
\begin{tabular}{|c|c c c|}
\hline
\bfseries Error level & \multicolumn{3}{|c|}{\bfseries Daily }\\
\hline
\hline
$\bm{\sigma}$ & $\bm{\beta}$ &$\bm{\gamma}$ &$\bm{\alpha}$\\
\hline
0\% & 0.01 & 0.12 & 0.88 \\
1\% & 0.94 & 6.84 & 53.09 \\
5\% & 3.67 & 23.59 & 154.27 \\
10\% & 7.36 & 45.10 & 225.45 \\
20 \%& 14.43 & 80.58 & 284.47 \\
30 \%& 21.38 & 95.99 & 291.84 \\
\hline
\bfseries Identifiable &  \cellcolor[HTML]{8EA9DB}Yes & No & No \\
\hline
\end{tabular}
\vspace{-0.20cm}

\caption{Cumulative Incidence Scenario 4 }
\label{Cumulative Incidence Scenario 4 FMSB}
\end{table}

\newpage
\newpage

\section{Supplemental Tables - Monte Carlo Results from \texttt{fmincon}}

\begin{table}[ht]
\centering
\caption{Prevalence Scenario 1}
\begin{tabular}{|c|c c c|c c c|c c c|}
\hline
\bfseries Error level  & \multicolumn{3}{|c|}{\bfseries Daily } & \multicolumn{3}{|c|}{\bfseries Weekly}& \multicolumn{3}{|c|}{\bfseries Monthly }\\
\hline
\hline
$\bm{\sigma}$ & $\bm{\beta}$ &$\bm{\gamma}$ &$\bm{\alpha}$ & $\bm{\beta}$ &$\bm{\gamma}$ &$\bm{\alpha}$ & $\bm{\beta}$ &$\bm{\gamma}$ &$\bm{\alpha}$\\
\hline
0 \%& 0.00 & 0.00 & 0.00  & 0.00 & 0.00 & 0.00 & 0.00 & 0.00 & 0.00 \\
1 \%& 0.05 & 0.18 & 0.04  & 0.19 & 0.84 & 0.11 & 0.46 & 2.16 & 0.24 \\
5 \% & 0.40 & 1.87 & 0.23  & 1.13 & 5.32 & 0.57 & 2.40 & 11.46 & 1.16 \\
10 \% & 0.91 & 4.07 & 0.72  & 2.39 & 10.99 & 1.21 & 5.28 & 24.59 & 2.33 \\
20\% & 2.47 & 9.38 & 2.79 & 5.99 & 22.86 & 3.07 & 17.08 & 59.54 & 4.55 \\
30 \%& 5.39 & 16.33 & 6.10  & 13.51 & 36.05 & 5.85 & 28.85 & 85.52 & 6.49 \\
\hline
\bfseries Identifiable  &\cellcolor[HTML]{8EA9DB} Yes & \cellcolor[HTML]{8EA9DB}Yes &\cellcolor[HTML]{8EA9DB} Yes & \cellcolor[HTML]{8EA9DB}Yes & No & \cellcolor[HTML]{8EA9DB}Yes & \cellcolor[HTML]{8EA9DB}Yes & No & \cellcolor[HTML]{8EA9DB}Yes\\
\hline
\end{tabular}
\vspace{-0.20cm}

\label{Prevalence Scenario 1 FMC}
\end{table}

\vspace{-0.30cm}
\begin{table}[H]
\centering
\caption{Prevalence Scenario 2}
\begin{tabular}{|c|c c c|c c c|}
\hline
\bfseries Error level & \multicolumn{3}{|c|}{\bfseries Daily } & \multicolumn{3}{|c|}{\bfseries Weekly}\\
\hline
\hline
$\bm{\sigma}$ & $\bm{\beta}$ &$\bm{\gamma}$ &$\bm{\alpha}$ & $\bm{\beta}$ &$\bm{\gamma}$ &$\bm{\alpha}$\\
\hline
0 \%& 0.01 & 0.01 & 0.00 &  0.02 & 0.02 & 0.00  \\
1 \%& 1.06 & 1.32 & 0.24 &	2.85 &	3.53 &	0.58\\
5 \%& 5.44 & 7.05 & 1.27 &  15.60 & 32.17 & 2.89  \\
10\% & 11.70 & 21.41 & 2.76 &  28.35 & 67.98 & 5.80   \\
20 \%& 24.41 & 73.83 & 7.25 &  45.01 & 95.99 & 12.26   \\
30 \%& 32.74 & 116.67 & 14.49 &  69.30 & 106.77 & 19.20  \\
\hline
\bfseries Identifiable  & No & No & \cellcolor[HTML]{8EA9DB}Yes & No & No & \cellcolor[HTML]{8EA9DB}Yes \\
\hline
\end{tabular}
\vspace{-0.20cm}

\label{Prevalence Scenario 2 FMC}
\end{table}

\vspace{-0.30cm}

\begin{table}[H]
\centering
\caption{Prevalence Scenario 3}
\begin{tabular}{|c|c c c|c c c|c c c|}
\hline
\bfseries Error level  & \multicolumn{3}{|c|}{\bfseries Daily } & \multicolumn{3}{|c|}{\bfseries Weekly }& \multicolumn{3}{|c|}{\bfseries Monthly}\\
\hline
\hline
$\bm{\sigma}$ & $\bm{\beta}$ &$\bm{\gamma}$ &$\bm{\alpha}$ & $\bm{\beta}$ &$\bm{\gamma}$ &$\bm{\alpha}$ & $\bm{\beta}$ &$\bm{\gamma}$ &$\bm{\alpha}$\\
\hline
0\% & 0.00 & 0.00 & 0.00  & 0.00 & 0.00 & 0.00  & 0.00 & 0.00 & 0.00 \\
1\% & 0.50 & 0.91 & 0.66  & 1.36 & 3.01 & 1.62  & 3.85 & 8.93 & 4.24 \\
5\%& 2.73 & 6.03 & 3.41  & 6.63 & 15.21 & 7.94  & 21.99 & 68.52 & 21.20 \\
10\% & 5.38 & 13.17 & 6.76  & 13.70 & 36.26 & 15.88  & 47.07 & 129.59 & 34.81 \\
20\% & 10.98 & 39.95 & 13.94 & 32.05 & 95.21 & 29.14  & 73.95 & 163.65 & 52.35 \\
30\% & 16.80 & 108.70 & 21.93  & 45.91 & 145.96 & 37.28  & 2851.12 & 173.97 & 68.07 \\
\hline
\bfseries Identifiable  & \cellcolor[HTML]{8EA9DB}Yes & No & \cellcolor[HTML]{8EA9DB}Yes & No & No & No  &  No & No & No\\
\hline
\end{tabular}
\vspace{-0.20cm}

\label{Prevalence Scenario 3 FMC}
\end{table}
\vspace{-0.30cm}

\begin{table}[H]
\centering
\caption{Prevalence Scenario 4}
\begin{tabular}{|c|c c c|}
\hline
\bfseries Error level  & \multicolumn{3}{|c|}{\bfseries Daily }\\
\hline
\hline
$\bm{\sigma}$ & $\bm{\beta}$ &$\bm{\gamma}$ &$\bm{\alpha}$\\
\hline
0\% & 2.26 & 3.93 & 5.46 \\
1\% & 7.46 & 13.18 & 18.50 \\
5\% & 19.16 & 38.24 & 45.39 \\
10\% & 25.85 & 60.22 & 57.45 \\
20\% & 36.36 & 103.55 & 76.31 \\
30\% & 45.01 & 133.93 & 90.46\\
\hline
\bfseries Identifiable & No & No & No \\
\hline
\end{tabular}
\vspace{-0.20cm}

\label{Prevalence Scenario 4 FMC}
\end{table}

\newpage
\begin{table}[H]
\centering
\begin{tabular}{|p{1cm}|c c c|c c c|c c c|}
\hline
\bfseries Error level & \multicolumn{3}{|c|}{\bfseries Daily} & \multicolumn{3}{|c|}{\bfseries Weekly}& \multicolumn{3}{|c|}{\bfseries Monthly }\\
\hline
\hline
$\bm{\sigma}$ & $\bm{\beta}$ &$\bm{\gamma}$ &$\bm{\alpha}$ & $\bm{\beta}$ &$\bm{\gamma}$ &$\bm{\alpha}$ & $\bm{\beta}$ &$\bm{\gamma}$ &$\bm{\alpha}$\\
\hline
0\% & 0.00& 0.00 & 0.00 &0.00 &0.00 & 0.00& 0.00& 0.00 & 0.00 \\
1\% & 0.01 & 0.03 & 0.02  & 0.07 & 0.52 & 0.19  & 0.20 & 1.87 & 0.51 \\
5\% & 0.20 & 1.65 & 0.48  & 0.49 & 4.90 & 1.35  & 0.97 & 10.49 & 2.71 \\
10\% & 0.45 & 3.89 & 1.01  & 0.97 & 10.20 & 2.69  & 1.96 & 23.65 & 5.49 \\
20\% & 1.10 & 8.42 & 1.97  & 2.01 & 22.71 & 5.13  & 3.72 & 61.78 & 10.80 \\
30\% & 2.12 & 13.89 & 2.65  & 3.16 & 39.27 & 7.08  & 5.69 & 99.19 & 16.78 \\
\hline
\bfseries Identifiable  & \cellcolor[HTML]{8EA9DB}Yes & \cellcolor[HTML]{8EA9DB}Yes & \cellcolor[HTML]{8EA9DB}Yes  & \cellcolor[HTML]{8EA9DB}Yes & No &\cellcolor[HTML]{8EA9DB} Yes  & \cellcolor[HTML]{8EA9DB}Yes & No & No \\
\hline
\end{tabular}
\vspace{-0.20cm}

\caption{Incidence Scenario 1}
\label{Incidence Scenario 1 FMC}
\end{table}

\vspace{-0.30cm}
\begin{table}[H]
\centering
\begin{tabular}{|c|c c c|c c c|c c c|}
\hline
\bfseries Error level & \multicolumn{3}{|c|}{\bfseries Daily} & \multicolumn{3}{|c|}{\bfseries Weekly}\\
\hline
\hline
$\bm{\sigma}$ & $\bm{\beta}$ &$\bm{\gamma}$ &$\bm{\alpha}$ & $\bm{\beta}$ &$\bm{\gamma}$ &$\bm{\alpha}$ \\
\hline
0\%& 0.00 & 0.00 & 0.00  & 0.00 & 0.00 & 0.00  \\
1\%& 0.13 & 0.24 & 0.28  & 0.30 & 0.60 & 0.66   \\
5\%& 0.56 & 1.08 & 1.26 & 1.70 & 3.31 & 3.71  \\
10\% & 1.22 & 2.30 & 2.73  & 3.39 & 6.45 & 7.32  \\
20\% & 2.78 & 5.01 & 6.11  & 6.55 & 11.57 & 13.58   \\
30\%& 5.20 & 8.51 & 10.83  & 9.50 & 16.26 & 18.81  \\
\hline
\bfseries Identifiable  & \cellcolor[HTML]{8EA9DB}Yes &\cellcolor[HTML]{8EA9DB} Yes & \cellcolor[HTML]{8EA9DB}Yes  & \cellcolor[HTML]{8EA9DB}Yes & \cellcolor[HTML]{8EA9DB}Yes &\cellcolor[HTML]{8EA9DB} Yes   \\
\hline
\end{tabular}
\vspace{-0.20cm}

\caption{Incidence Scenario 2}
\label{Incidence Scenario 2 FMC}
\end{table}
\vspace{-0.30cm}

\begin{table}[H]
\centering
\begin{tabular}{|c|c c c|c c c|c c c|}
\hline
\bfseries Error level  & \multicolumn{3}{|c|}{\bfseries Daily } & \multicolumn{3}{|c|}{\bfseries Weekly }& \multicolumn{3}{|c|}{\bfseries Monthly }\\
\hline
\hline
$\bm{\sigma}$ & $\bm{\beta}$ &$\bm{\gamma}$ &$\bm{\alpha}$ & $\bm{\beta}$ &$\bm{\gamma}$ &$\bm{\alpha}$ & $\bm{\beta}$ &$\bm{\gamma}$ &$\bm{\alpha}$\\
\hline
0\% & 0.00 & 0.00 & 0.00 & 0.00 & 0.02 & 0.00 & 0.00 & 0.00 & 0.00 \\
1\% & 0.33 & 1.33 & 0.55 & 1.44 & 5.63 & 1.92 & 5.61 & 30.30 & 7.85 \\
5\% & 1.79 & 7.20 & 2.76 & 7.14 & 38.76 & 9.66 & 28.82 & 154.69 & 30.21 \\
10 \%& 3.67 & 14.51 & 5.42 & 12.75 & 79.46 & 17.71 & 46.28 & 188.07 & 47.96 \\
20\% & 8.47 & 30.64 & 11.08 & 21.02 & 111.28 & 33.81 & 68.55 & 182.10 & 85.63 \\
30\% & 15.19 & 45.31 & 19.21 &  28.01 & 119.14 & 53.23 & 93.48 & 154.56 & 125.72 \\
\hline
\bfseries Identifiable  &  \cellcolor[HTML]{8EA9DB}Yes & No &  \cellcolor[HTML]{8EA9DB}Yes &   \cellcolor[HTML]{8EA9DB}Yes & No & No & No & No & No \\
\hline
\end{tabular}
\vspace{-0.20cm}

\caption{Incidence Scenario 3 }
\label{Incidence Scenario 3 FMC}
\end{table}
\vspace{-0.30cm}

\begin{table}[H]
\centering
\begin{tabular}{|c|c c c|}
\hline
\bfseries Error level  & \multicolumn{3}{|c|}{\bfseries Daily }\\
\hline
\hline
$\bm{\sigma}$ & $\bm{\beta}$ &$\bm{\gamma}$ &$\bm{\alpha}$\\
\hline
0\% & 0.01 & 0.00 & 0.00 \\
1\% & 0.52	& 2.25 &	14.04 \\
5\% & 2.57	& 10.02 &	58.51 \\
10\% & 5.24	& 18.75	& 99.72 \\
20\% & 10.85 & 36.77 &	155.21 \\
30\% & 16.92 &	49.96 &	183.86 \\
\hline
\bfseries Identifiable & \cellcolor[HTML]{8EA9DB} Yes & No & No \\
\hline
\end{tabular}
\vspace{-0.20cm}

\caption{Incidence  Scenario 4 }
\label{Incidence Scenario 4 FMC}
\end{table}
\vspace{-0.30cm}

\begin{table}[H]
\centering
\begin{tabular}{|c|c c c|c c c|c c c|}
\hline
\bfseries Error level  & \multicolumn{3}{|c|}{\bfseries Daily } & \multicolumn{3}{|c|}{\bfseries Weekly }& \multicolumn{3}{|c|}{\bfseries Monthly }\\
\hline
\hline
$\bm{\sigma}$ & $\bm{\beta}$ &$\bm{\gamma}$ &$\bm{\alpha}$ & $\bm{\beta}$ &$\bm{\gamma}$ &$\bm{\alpha}$ & $\bm{\beta}$ &$\bm{\gamma}$ &$\bm{\alpha}$\\
\hline
     0\% & 0.00 & 0.00 & 0.00 & 0.00 & 0.00 & 0.00 & 0.00 & 0.00 & 0.00 \\
        1\% & 0.43 & 1.54 & 0.63 & 1.48 & 6.12 & 1.78 & 4.90 & 29.22 & 5.54 \\
        5\% & 2.69 & 10.82 & 3.36 & 7.90 & 54.30 & 9.24 & 18.30 & 131.12 & 20.40 \\
        10\% & 6.43 & 28.93 & 6.87 & 14.64 & 103.96 & 17.28 & 29.64 & 162.19 & 39.33 \\
        20\% & 17.75 & 61.80 & 16.55 & 26.46 & 138.19 & 40.81 & 47.06 & 170.81 & 72.01 \\
        30\% & 29.35 & 76.12 & 48.68 & 35.54 & 135.64 & 64.31 & 1839.19 & 157.51 & 98.54 \\
\hline
\bfseries Identifiable  &  \cellcolor[HTML]{8EA9DB}Yes& No & No & No & No & No & No & No & No \\
\hline
\end{tabular}
\vspace{-0.20cm}

\caption{Cumulative Incidence Scenario 1}
\label{fmc:Cumulative Incidence Scenario 1}
\end{table}

\vspace{-0.30cm}
\begin{table}[H]
\centering
\begin{tabular}{|c|c c c|c c c|}
\hline
\bfseries Error level  & \multicolumn{3}{|c|}{\bfseries Daily } & \multicolumn{3}{|c|}{\bfseries Weekly }\\
\hline
\hline
$\bm{\sigma}$ & $\bm{\beta}$ &$\bm{\gamma}$ &$\bm{\alpha}$ & $\bm{\beta}$ &$\bm{\gamma}$ &$\bm{\alpha}$ \\
\hline
  0\% & 1.23 & 13.89 & 105.71 & 0.12 & 1.89 & 9.89 \\
        1\% & 0.99 & 8.07 & 61.47 & 5.90 & 52.39 & 158.97 \\
        5\% & 3.66 & 21.79 & 139.17 & 20.27 & 143.20 & 226.75 \\
        10\% & 7.28 & 35.27 & 182.15 & 32.51 & 174.51 & 305.53 \\
        20\% & 14.12 & 53.18 & 179.58 & 173.07 & 187.73 & 389.63 \\
        30\% & 20.60 & 67.95 & 155.22 & 1977.03 & 188.83 & 413.24 \\
\hline
\bfseries Identifiable  &  \cellcolor[HTML]{8EA9DB}Yes & No & No & No & No & No \\
\hline
\end{tabular}
\vspace{-0.20cm}

\caption{Cumulative Incidence Scenario 2 }
\label{fmincon:Cumulative Incidence Scenario 2}
\end{table}

\vspace{-0.30cm}
\begin{table}[H]
\centering
\begin{tabular}{|c|c c c|c c c|c c c|}
\hline
\bfseries Error level  & \multicolumn{3}{|c|}{\bfseries Daily } & \multicolumn{3}{|c|}{\bfseries Weekly }& \multicolumn{3}{|c|}{\bfseries Monthly }\\
\hline
\hline
$\bm{\sigma}$ & $\bm{\beta}$ &$\bm{\gamma}$ &$\bm{\alpha}$ & $\bm{\beta}$ &$\bm{\gamma}$ &$\bm{\alpha}$ & $\bm{\beta}$ &$\bm{\gamma}$ &$\bm{\alpha}$\\
\hline
0\% & 0.00 & 0.00 & 0.00 & 0.00 & 0.00 & 0.00 & 0.00 & 0.00 & 0.00 \\
1 \%& 0.50 & 2.04 & 0.84 &  1.62 & 8.62 & 2.43 &  6.26 & 44.92 & 7.50 \\
5 \%& 3.06 & 17.52 & 4.55 &  8.55 & 76.71 & 13.34 &  18.25 & 170.24 & 36.76 \\
10\% & 6.65 & 43.63 & 9.82 & 13.67 & 135.67 & 30.67 &  23.80 & 201.17 & 60.30 \\
20 \%& 13.96 & 75.08 & 25.29 & 20.52 & 169.93 & 55.42 &  38.55 & 206.35 & 92.92 \\
30 \%& 23.18 & 82.04 & 43.28 &  28.47 & 178.60 & 74.76 &  1009.17 & 201.16 & 143.25 \\
\hline
\bfseries Identifiable  &  \cellcolor[HTML]{8EA9DB}Yes & No & No &  \cellcolor[HTML]{8EA9DB}Yes & No & No & No & No & No \\
\hline
\end{tabular}
\vspace{-0.20cm}

\caption{Cumulative Incidence Scenario 3 }
\label{fmincon:Cumulative Incidence Scenario 3}
\end{table}

\vspace{-0.30cm}

\begin{table}[H]
\centering
\begin{tabular}{|c|c c c|}
\hline
\bfseries Error level & \multicolumn{3}{|c|}{\bfseries Daily }\\
\hline
\hline
$\bm{\sigma}$ & $\bm{\beta}$ &$\bm{\gamma}$ &$\bm{\alpha}$\\
\hline
0\% & 0.02 & 0.30 & 2.18 \\
1\% & 1.01 & 8.24 & 62.80 \\
5\% & 3.72 & 23.96 & 151.22 \\
10\% & 7.44 & 45.34 & 221.31 \\
20 \%& 14.54 & 80.70 & 284.35 \\
30 \%& 21.53 & 96.19 & 296.40 \\
\hline
\bfseries Identifiable &  \cellcolor[HTML]{8EA9DB}Yes & No & No \\
\hline
\end{tabular}
\vspace{-0.20cm}

\caption{Cumulative Incidence Scenario 4 }
\label{fmcon:Cumulative Incidence Scenario 4}
\end{table}

\newpage

\section{Supplemental Tables - Correlation Matrix Results}

The following tables contain the correlation coefficients for the different scenarios and data types, rounded to two decimal places. If the correlation coefficient is above 0.9 for any pair of parameters, the parameter set is `not identifiable' for the scenario and data type.


\begin{table}[H]
\centering
\begin{tabular}{|l|ccc|cc|ccc|c|}
\hline
{\bfseries Scenario} & \multicolumn{3}{c|}{\bfseries 1} & \multicolumn{2}{c|}{ \bfseries 2}&\multicolumn{3}{c|}{ \bfseries  3}& \bfseries   4  \\
{\bfseries Sampling}    &\bfseries D & \bfseries W &\bfseries   M & \bfseries D &\bfseries  W & \bfseries  D &\bfseries W &\bfseries M &\bfseries D \\
     \hline
     \hline
$\beta,\gamma$ correlation   & -0.98                         & -0.98                         & -0.98                         & -0.99                         & -0.99                         & \cellcolor[HTML]{8EA9DB}-0.86 & \cellcolor[HTML]{8EA9DB}-0.87 & -0.95                         & -1.00                      \\
$\beta,\alpha$ correlation    & \cellcolor[HTML]{8EA9DB}-0.66 & \cellcolor[HTML]{8EA9DB}-0.67 & \cellcolor[HTML]{8EA9DB}-0.68 & \cellcolor[HTML]{8EA9DB}0.01 & \cellcolor[HTML]{8EA9DB}0.10 & \cellcolor[HTML]{8EA9DB}0.90 & \cellcolor[HTML]{8EA9DB}0.89 & 0.94                         & -0.97                        \\
$\gamma,\alpha$ correlation    & \cellcolor[HTML]{8EA9DB}0.54 & \cellcolor[HTML]{8EA9DB}0.54 & \cellcolor[HTML]{8EA9DB}0.56 & \cellcolor[HTML]{8EA9DB}-0.01 & \cellcolor[HTML]{8EA9DB}-0.10 & \cellcolor[HTML]{8EA9DB}-0.55 & \cellcolor[HTML]{8EA9DB}-0.57 & \cellcolor[HTML]{8EA9DB}-0.79 & 0.99                        \\
\hline

\bfseries Identifiable?   & No                           & No                           & No                           & No                           & No                           & \cellcolor[HTML]{8EA9DB}Yes  & \cellcolor[HTML]{8EA9DB}Yes  & No                           & No                          \\
\bfseries MCMC Results & \cellcolor[HTML]{8EA9DB} $\beta$, $\gamma$, $\alpha$   & $\beta$, $\alpha$                            & $\beta$, $\alpha$                            & $\alpha$                            & $\alpha$                            & $\beta$, $\alpha$                            & -                            & -                            & -  \\    \hline
\end{tabular}
\caption{Correlation matrix identifiability results for Prevalence data. For comparison, we include the identifiable parameters from MCMC results in the final row.}
\label{CMPrev}
\end{table}

\begin{table}[H]
\centering
\begin{tabular}{|l|ccc|cc|ccc|c|}
\hline
{\bfseries Scenario} & \multicolumn{3}{c|}{\bfseries 1} & \multicolumn{2}{c|}{ \bfseries 2}&\multicolumn{3}{c|}{ \bfseries  3}& \bfseries   4  \\
{\bfseries Sampling}    &\bfseries D & \bfseries W &\bfseries   M & \bfseries D &\bfseries  W & \bfseries  D &\bfseries W &\bfseries M &\bfseries D \\
     \hline
     \hline
$\beta,\gamma$ correlation             & \cellcolor[HTML]{8EA9DB}-0.84                        & \cellcolor[HTML]{8EA9DB}-0.84 & \cellcolor[HTML]{8EA9DB}-0.82 & -1.00                                                & -1.00                                                & \cellcolor[HTML]{8EA9DB}-0.90 & -0.90                         & -0.97                         & \cellcolor[HTML]{8EA9DB}-0.32 \\
$\beta,\alpha$ correlation             & \cellcolor[HTML]{8EA9DB}-0.14                        & \cellcolor[HTML]{8EA9DB}-0.14 & \cellcolor[HTML]{8EA9DB}-0.11 & 1.00                                                & 1.00                                                & \cellcolor[HTML]{8EA9DB}0.74 & \cellcolor[HTML]{8EA9DB}0.74 & \cellcolor[HTML]{8EA9DB}0.83 & \cellcolor[HTML]{8EA9DB}0.20 \\
$\gamma,\alpha$ correlation            & \cellcolor[HTML]{8EA9DB}0.64                        & \cellcolor[HTML]{8EA9DB}0.64 & \cellcolor[HTML]{8EA9DB}0.64 & -0.99                                                & -0.99                                                & \cellcolor[HTML]{8EA9DB}-0.38 & \cellcolor[HTML]{8EA9DB}-0.39 & \cellcolor[HTML]{8EA9DB}-0.69 & \cellcolor[HTML]{8EA9DB}0.86 \\
     \hline
\bfseries Identifiable? & \cellcolor[HTML]{8EA9DB}Yes                         & \cellcolor[HTML]{8EA9DB}Yes  & \cellcolor[HTML]{8EA9DB}Yes  & No                                                  & No                                                  & \cellcolor[HTML]{8EA9DB}Yes  & No                           & No                           & \cellcolor[HTML]{8EA9DB}Yes  \\
\bfseries MCMC Results  & \cellcolor[HTML]{8EA9DB}$\beta$, $\gamma$, $\alpha$ & $\beta$, $\alpha$            & $\beta$                      & \cellcolor[HTML]{8EA9DB}$\beta$, $\gamma$, $\alpha$ & \cellcolor[HTML]{8EA9DB}$\beta$, $\gamma$, $\alpha$ & $\beta$                      & -                            & -                            & -         \\
     \hline
\end{tabular}
\caption{Correlation matrix identifiability results for Incidence data. For comparison, we include the identifiable parameters from MCMC results in the final row.}
\label{CMIncid}
\end{table}

\begin{table}[H]
\centering
\begin{tabular}{|l|ccc|cc|ccc|c|}
\hline
{\bfseries Scenario} & \multicolumn{3}{c|}{\bfseries 1} & \multicolumn{2}{c|}{ \bfseries 2}&\multicolumn{3}{c|}{ \bfseries  3}& \bfseries   4  \\
{\bfseries Sampling}    &\bfseries D & \bfseries W &\bfseries   M & \bfseries D &\bfseries  W & \bfseries  D &\bfseries W &\bfseries M &\bfseries D \\
     \hline
     \hline
$\beta,\gamma$ correlation             & -0.97                         & -0.97                         & -0.99    & \cellcolor[HTML]{8EA9DB}0.56 & \cellcolor[HTML]{8EA9DB}-0.68 & -0.92                         & -0.92                         & -0.98                         & \cellcolor[HTML]{8EA9DB}0.58 \\
$\beta,\alpha$ correlation             & 0.92                         & 0.92                         & 0.97    & \cellcolor[HTML]{8EA9DB}0.73 & \cellcolor[HTML]{8EA9DB}-0.59 & \cellcolor[HTML]{8EA9DB}0.33 & \cellcolor[HTML]{8EA9DB}0.38 & \cellcolor[HTML]{8EA9DB}0.75 & \cellcolor[HTML]{8EA9DB}0.75 \\
$\gamma,\alpha$ correlation            & \cellcolor[HTML]{8EA9DB}-0.81 & \cellcolor[HTML]{8EA9DB}-0.81 & -0.93    & 0.97                         & 0.99                         & \cellcolor[HTML]{8EA9DB}0.06 & \cellcolor[HTML]{8EA9DB}0.00 & \cellcolor[HTML]{8EA9DB}-0.61 & 0.97                         \\
\hline
\bfseries Identifiable? & No                           & No                           & No      & No                           & No                           & No                           & No                           & No                           & No                           \\
\bfseries MCMC Results  & -                            & -                            & -       & -                            & -                            & -  & - & -    & -    \\
     \hline
\end{tabular}
\caption{Correlation matrix identifiability results for Cumulative Incidence data. For comparison, we include the identifiable parameters from MCMC results in the final row.}
\label{CMCuml}
\end{table}

The following tables report the percent of identifiable parameters from MCMC iterates. In three scenarios, we omit parameter estimates which result in a noninvertible Fischer Information Matrix. This happens for monthly Prevalence data under Scenario 3 (984 parameter sets omitted), weekly Cumulative Incidence data under Scenario 2 (4 parameter sets omitted), and monthly Cumulative Incidence data under Scenario 2 (641 parameter sets omitted). The corresponding results are indicated with an asterisk in the tables below.

\begin{table}[H]
\centering
\begin{tabular}{|l|ccc|cc|ccc|c|}
\hline
{\bfseries Scenario} & \multicolumn{3}{c|}{\bfseries 1} & \multicolumn{2}{c|}{ \bfseries 2}&\multicolumn{3}{c|}{ \bfseries  3}& \bfseries   4  \\
{\bfseries Sampling}    &\bfseries D & \bfseries W &\bfseries   M & \bfseries D &\bfseries  W & \bfseries  D &\bfseries W &\bfseries M &\bfseries D \\
     \hline
     \hline
\bfseries 0\% Error Estimates  & 0\%   & 0\%   & 0\%   & 0\%       & 0\%       & 100\%  & 100\%  & 0\% & 0\% \\
\bfseries  30\% Error Estimates & 0\%   & 0\%   & 0\%   & 0\%       & 0.33\%       & 18\%   & 10\%   & 5.2*\% & 1.5\% \\
     \hline
\bfseries True Parameter Values      & No    & No    & No    & No        & No        & Yes    & Yes    & No  & No  \\
     \hline
\end{tabular}
\caption{Correlation matrix identifiability rates for MCMC parameter estimates, using Prevalence data. For comparison, we include results for true parameters in the last row. The asterisk indicates a case where parameter estimates were omitted due to noninvertible Fisher Matrices.}
\label{MCMCPrev}
\end{table}

\begin{table}[H]
\centering
\begin{tabular}{|l|ccc|cc|ccc|c|}
\hline
{\bfseries Scenario} & \multicolumn{3}{c|}{\bfseries 1} & \multicolumn{2}{c|}{ \bfseries 2}&\multicolumn{3}{c|}{ \bfseries  3}& \bfseries   4  \\
{\bfseries Sampling}    &\bfseries D & \bfseries W &\bfseries   M & \bfseries D &\bfseries  W & \bfseries  D &\bfseries W &\bfseries M &\bfseries D \\
     \hline
     \hline
     \bfseries 0\% Error Estimates  & 100\% & 100\% & 100\% & 0\%       & 0\%       & 100\%  & 0\%   & 0\%  & 100\% \\
\bfseries 30\% Error Estimates & 100\% & 100\% & 96\%  & 0\%       & 0\%       & 86\%   & 45\%  & 6.2\% & 63\% \\  \hline
\bfseries True parameters      & Yes   & Yes   & Yes   & No        & No        & Yes    & No    & No   & Yes \\

     \hline
\end{tabular}
\caption{Correlation matrix identifiability rates for MCMC parameter estimates, using Incidence data. For comparison, we include results for true parameters in the last row.}
\label{MCMCIncid}
\end{table}

\begin{table}[H]
\centering
\begin{tabular}{|l|ccc|cc|ccc|c|}
\hline
{\bfseries Scenario} & \multicolumn{3}{c|}{\bfseries 1} & \multicolumn{2}{c|}{ \bfseries 2}&\multicolumn{3}{c|}{ \bfseries  3}& \bfseries   4  \\
{\bfseries Sampling}    &\bfseries D & \bfseries W &\bfseries   M & \bfseries D &\bfseries  W & \bfseries  D &\bfseries W &\bfseries M &\bfseries D \\
     \hline
     \hline
     \bfseries 0\% Error Estimates  & 0\%   & 0\%   & 0\%   & 100\%     & 100\%     & 0\%   & 0\%   & 0\%   & 100\% \\
\bfseries 30\% Error Estimates & 66\%  & 34\%  & 13\%  & 17\%      & 3*\%      & 38\%  & 15\%  & 1.8*\%  & 25\%  \\  \hline
\bfseries True parameters      & No    & No    & No    & No        & No        & No    & No    & No    & No  \\
     \hline
\end{tabular}
\caption{Correlation matrix identifiability rates for MCMC parameter estimates, using Cumulative Incidence data. For comparison, we include results for true parameters in the last row. The asterisks indicate cases where parameter estimates were omitted due to noninvertible Fisher Matrices.}
\label{MCMCCuml}
\end{table}

\end{document}